\documentclass[aps,prb,amsmath,amssymb,reprint,superscriptaddress,showkeys]{revtex4-2}

\usepackage{multirow}
\usepackage{graphicx}
\usepackage{amsmath}
\usepackage{textcomp}
\usepackage{pifont}
\usepackage{tikz}
\usepackage{blindtext}

\usepackage[colorlinks]{hyperref}

\DeclareUnicodeCharacter{202F}{\,}
\usepackage{ulem}

\begin{document}

\title{Microstructural analysis of GaN films grown on (1 0 0) MgF\texorpdfstring{$_{2}$}{2} substrate by 4D nanobeam diffraction and energy-dispersive X-ray spectrometry}

\author{Tobias Niemeyer}
\affiliation{4th Institute of Physics \textendash{} Solids and Nanostructures, University of Goettingen, Friedrich-Hund-Platz 1, 37077 G\"ottingen, Germany}
%\ead{tobias.westphal@uni-goettingen.de}
\author{Kevin Meyer}
\affiliation{Institute of Energy Research and Physical Technologies IEPT, Clausthal University of Technology, Leibnizstrasse 4, 38678 Clausthal-Zellerfeld, Germany}
\author{Christoph Flathmann}
\affiliation{4th Institute of Physics \textendash{} Solids and Nanostructures, University of Goettingen, Friedrich-Hund-Platz 1, 37077 G\"ottingen, Germany}
\author{Tobias Meyer}
\affiliation{Institute of Materials Physics, University of Goettingen, Friedrich-Hund-Platz 1, 37077 G\"ottingen, Germany}
\author{Daniel M. Schaadt}
\affiliation{Institute of Energy Research and Physical Technologies IEPT, Clausthal University of Technology, Leibnizstrasse 4, 38678 Clausthal-Zellerfeld, Germany}
\author{Michael Seibt}%\texorpdfstring{\corref{cor1}}{*}}
\email{michael.seibt@uni-goettingen.de}
\affiliation{4th Institute of Physics \textendash{} Solids and Nanostructures, University of Goettingen, Friedrich-Hund-Platz 1, 37077 G\"ottingen, Germany}

\begin{abstract}
The use of highly efficient and solarblind GaN photocathodes as part of multichannel plate UV detectors for applications in astronomy would strongly benefit from the direct growth of GaN on typical window materials with high transmission down to the deep UV range. GaN growth on MgF$_2$ substrates by plasma-assisted molecular beam epitaxy has recently been demonstrated. Here, we report an extensive scanning transmission electron microscopy study of the thin film microstructure for growth at 525\,$^{\circ}$C and 650\,$^{\circ}$C on (100) MgF$_2$. These results are systematically supported by X-ray diffraction reciprocal space maps. For both growth temperatures predominant cubic (111), (115) and (110) GaN is found with no preferred nucleation on the substrate and typical grain sizes of 100-200~nm. All observed orientations can be understood as the result of first and second order twins on different $\{111\}$ planes related to the underlying substrate. The higher growth temperature shows a strongly increased twin density along with a higher surface roughness. Furthermore, grains with cubic (110) GaN growth show a reduced density and a reduced size of about 20~nm. Furthermore, in-diffusion of Mg and F into the GaN is observed, which is accompanied by the formation of cavities in the MgF$_2$ directly at the interface.  
\end{abstract}


\maketitle

\section{Introduction}\label{sec01}

Group III-nitrides are promising materials for a plethora of device applications e.g. in electronics, photonics, photovoltaics, thermoelectrics and spintronics \cite{Zhou2017}. Here we focus on III-nitrides as solarblind photodetectors \cite{Parish1999,Cai2021} and in particular on high-quality GaN photocathodes as part of UV microchannel plate detectors for application in astronomy as conceptionalized previously \cite{Siegmund2004}. For such applications, GaN has to be combined with window materials, which are highly transparent in the deep UV. Typical substrate materials such as sapphire, silicon, GaAs, ZnO, 6H-SiC do not meet this requirement, while research on III-nitride growth on suitable window materials has just recently been initiated \cite{Meyer2020}. The authors studied MgF$_2$, which fulfills the optical constraints and has successfully been used as substrate for metal oxides such as TiO$_2$~\cite{Xie2016} and VO$_2$~\cite{Zhou2016}. In fact, epitaxial growth of mainly cubic GaN (c-GaN) in $\langle$111$\rangle$ and $\langle$110$\rangle$ orientation has been reported recently for Ga-rich growth conditions in plasma-assisted molecular beam epitaxy (PAMBE)~\cite{Meyer2020} as studied by X-ray diffraction (XRD). Further studies~\cite{Meyer2020a,Meyer2020a-corr} show that a higher growth temperature (650\,$^{\circ}$C) lead to substantial in-diffusion of Mg and F as evidenced by Secondary Ion Mass Spectrometry (SIMS). Both impurities are related to p-type doping of GaN \cite{Lyons2021}, which is traditionally used for the case of Mg~\cite{Amano1989,Nakamura1992} and a matter of active research for F, where the interstitial species is thought to be related to shallow acceptor states~\cite{Janotti2009,Liu2021}.

MgF$_2$ has a rutile crystal structure with $a=4.62$~\AA~and $c=3.05$~\AA~as lattice parameters. Hence, growth on (100) MgF$_2$ implies a twofold symmetry of the substrate as a growth template. Zincblende GaN (c-GaN) has a face centered cubic crystal structure with $a=4.52$~\AA~as lattice parameter and wurtzite GaN (h-GaN) with a hexagonal crystal structure has $a=3.19$~\AA~and $c=5.19$~\AA~as lattice parameters.

In this paper, we perform a detailed analysis of the GaN thin film microstructure obtained by plasma assisted molecular beam epitaxy (PAMBE) growth at 525\,$^{\circ}$C and 650\,$^{\circ}$C. We use (scanning) transmission electron microscopy, (S)TEM, imaging and analysis techniques to measure the spatial distribution of different GaN orientations in order to get insight into their nucleation and growth on (100) MgF$_2$ substrates. For this purpose, nanobeam electron diffraction (NBED)~\cite{Beche2009} is used in 4D-STEM mode, where NBED patterns are recorded for each beam position during the acquisition of a STEM image (see~\cite{Ophus2019} for a recent review). This extremely powerful approach allows for real space mapping of electric~\cite{Mueller2014} and magnetic~\cite{Krajnak2016} fields, materials domains~\cite{Brunetti2011}  as well as strain~\cite{Baumann2014,Ozdol2015} and orientation~\cite{Kobler2013} mapping of crystalline materials. Clearly, for the issues of this work, the latter two are most important, especially their ability of virtual dark-field imaging. These techniques are fully compatible with chemical analysis using energy dispersive X-ray spectrometry, EDX, which we use to monitor possible in-diffusion of Mg and F from MgF$_2$ substrates. To give a complete picture the 4D-STEM results are accompanied with high-resolution XRD (HRXRD) reciprocal space maps (RSM).

After giving experimental details in Sec.~\ref{sec02}, we start with the overall microstructure of GaN layers including surface roughness, cavity formation in the MgF$_2$ substrates at higher growth temperature, and a first observation of planar defects  (Sec.~\ref{sec031}). Subsequently, we present results of chemical analysis by means of EDX indicating an in-diffusion of Mg and F from the substrate at the higher growth temperature (Sec.~\ref{sec032}). Sec.~\ref{sec033} presents extensive NBED data obtained in 4D-STEM mode, where we show that GaN layers have grown in the cubic phase in five different orientations, which are (higher order) twins of each other (Sec.~\ref{sec0331}). With the identified c-GaN orientations a full overview of the experimental 4D-STEM data in combination with calculated diffraction patterns for both cross-section orientations along MgF$_2$ [010] and [001] is shown yielding an unambiguous and consistent description of all data (Sec.~\ref{sec0332}). Based on the observed orientations, a set of masks for both view directions is achieved to create virtual dark field (vDF) images for the individual orientations. Superimposing these vDFs gives an orientation map in real space for each growth temperature and lamella orientation showing the different positions of the different c-GaN orientations within the GaN layer (Sec.~\ref{sec0333}). These orientation maps show that there is no preferred nucleation of one orientation. The high density of twin boundaries at the higher growth temperature leads to additional observed spots from the second order Laue zone which is explained in Sec.~\ref{sec0333} of the microstructural analysis. Finally, HRXRD RSM results are shown supporting the findings from the 4D-STEM data (Sec.~\ref{sec034}).

\begin{figure*}
\centering
\includegraphics[width=.98\textwidth]{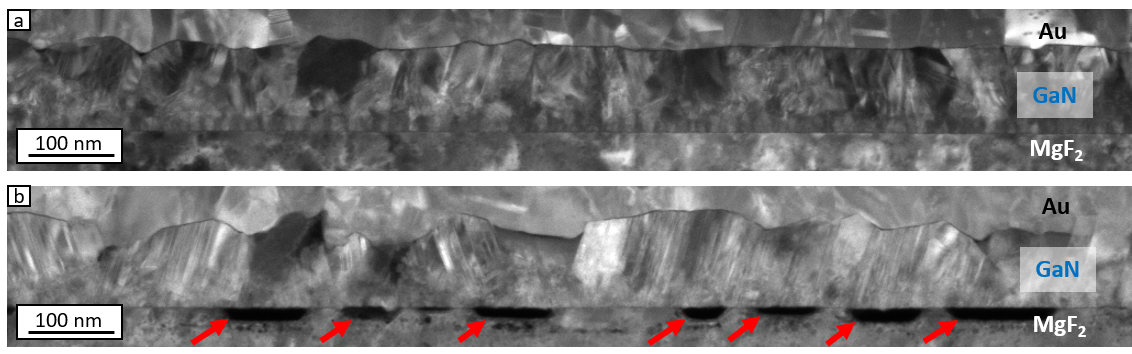}
\caption{ADF-STEM images in cross-section geometry of the GaN films on the MgF$_2$ substrate. (a) grown at 525\,$^{\circ}$C and (b) grown at 650\,$^{\circ}$C. In (b) hole formation at the interface between GaN and MgF$_2$ is visible and the surface of the GaN layer has a higher roughness in comparison with (a). For both samples the GaN layer is on average close to 100~nm thickness. The Au layer is thermally evaporated to reduce charging during FIB lamella preparation. In both images the zone axis of the MgF$_2$ substrate is [001].}
\label{Figure1}
\end{figure*}

\section{Material and Methods}\label{sec02}
The growth of  GaN layers has been performed by PAMBE on a (100)~MgF$_{2}$ substrate as described in detail elsewhere~\cite{Meyer2020a}. Briefly, PAMBE was used with reactive nitrogen from a radio-frequency plasma source with a gas flow of 0.2~sccm in order to establish Ga-rich growth conditions. For this study, samples were grown at 525\,$^{\circ}$C and 650\,$^{\circ}$C on (100) MgF$_2$ substrates. The low temperature as well as the Ga-rich conditions are selected to aim for cubic growth of GaN, reducing surface roughness and less damaging the MgF$_2$ substrate \cite{Meyer2020a}.

For TEM lamella preparation, conventional focused ion beam (FIB) procedure in cross-section geometry has been conducted using an FEI Helios G4 Dual Beam FIB. Final thinning of the lamellae has been done at 5~kV accelerating voltage of the ion beam. To avoid charging during the lamella preparation by FIB, a Au layer has been thermally evaporated prior to the preparation. For both growth temperatures cross-section lamellae perpendicular to the [010] and [001] orientation of the MgF$_2$ substrate have been prepared. These orientations have also been used as the zone axis (ZA) at the TEM experiments.

For the TEM investigations an image-corrected FEI Titan 80-300 ETEM G2 operated at 80~kV and 300~kV has been used. The energy-dispersive X-ray (EDX) data acquisition has been done with an Oxford Instruments X-Max 80~mm$^{2}$ detector and analyzed with HyperSpy \cite{HyperSpy}. The probe size during STEM is approximately 1.3~\AA. For the 4D-STEM acquisition a semi-convergence angle of 0.7~mrad is chosen leading to a probe size of approximately 2~nm. The NBED patterns have been recorded with an UltraScan 1000XP camera binned to 256$\times$256 pixels or 512$\times$512 pixels using a self-written DigitalMicrograph plugin to control beam position and read out of the camera \cite{Meyer2021}. Beam currents were tuned to 42~pA for imaging and to 100-150~pA for EDX.

The reciprocal space maps (RSM) were recorded with a Bruker D8 Discover HRXRD using a 2-bounce monochromator and a 0.2~mm slit in the primary side. The RSMs were built by rocking curves with a scan rate of 0.02\,$^{\circ}$.

\section{Results and Discussion}\label{sec03}
This section describes experimental results and their analysis and is organized such that a direct comparison between samples grown at 525\,$^{\circ}$C and 650\,$^{\circ}$C and the lamellae with the two different orientations is supported.

\subsection{Overall structural properties}\label{sec031}

The structural investigation starts with an annular dark field (ADF)-STEM overview of the GaN layer of both growth temperatures, shown in Fig.~\ref{Figure1}. By comparing the GaN layer of the samples at both temperatures, the higher roughness of the sample grown at 650\,$^{\circ}$C gets evident. This is in a good agreement with the AFM measurement in \cite{Meyer2020a} indicating a smaller roughness for the lower temperature. In addition, hole formation at the interface between GaN and MgF$_2$ for the 650\,$^{\circ}$C sample is visible in Fig.~\ref{Figure1}~(b). A closer look shows that the holes are extended into the MgF$_2$ substrate and not into the GaN layer. As there is no hole formation at the 525\,$^{\circ}$C sample visible which is apart from the temperature prepared in the same way, it is likely that the formation takes places during growth.

\subsection{Chemical analysis at substrate/film interfaces}\label{sec032}
\begin{figure}[ht]
\centering
\includegraphics[width=.9\columnwidth]{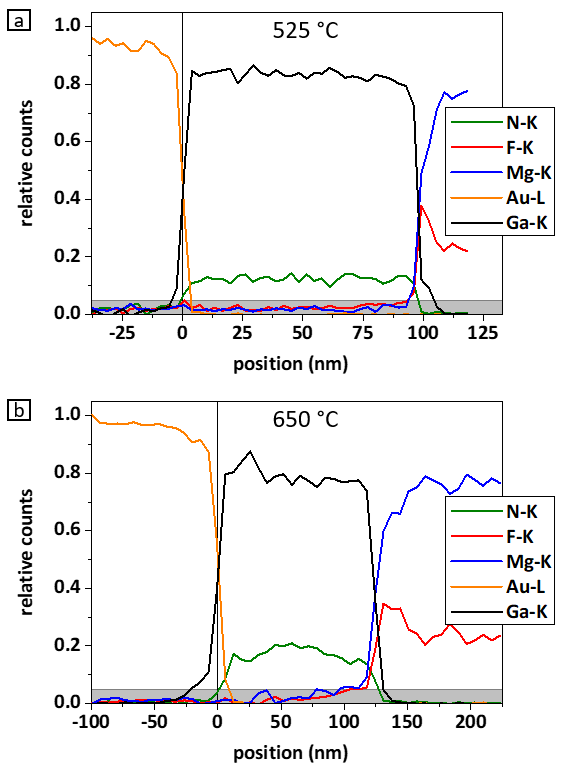}
\caption{EDX line profiles averaged over 5 lines scanned from the Au layer down to the substrate. The origin of the length scale is set to the point where the Au signal decreases to 0.5 relative counts. For both profiles the background is subtracted for each line individually. In (a) for the EDX profile of the 525\,$^{\circ}$C sample a sharp transition at both interfaces of the GaN layer is visible. For the 650\,$^{\circ}$C sample (b) a decrease of the Mg and F signal from the interface into the GaN layer is observed. The apparent broadening at the interface between Au and GaN stems from the averaging at the rough surface. The gray region at the bottom indicates the uncertainty due to the background subtraction.}
\label{figure2}
\end{figure}

According to recently published SIMS data~\cite{Meyer2020a}, PAMBE growth of GaN on MgF$_2$ leads to transport of Mg and F from the substrate into the GaN at higher growth temperatures. Since both impurities are related to p-type doping and bearing in mind the targeted use of GaN as a photocathode, this is an important issue. Fig.~\ref{figure2} shows qualitative concentration profiles of Ga, N, Mg and F obtained from the respective K-lines averaged over 5 scans at adjacent but non-overlapping lateral positions.

The measured EDX signals show the expected layer structure of Au, GaN and MgF$_2$ indicated by the Au, Ga, N, Mg and F signal. The background of the EDX signal is related to the material which is observed and is getting larger the heavier the elements. This leads to a different background for each of the three layers, where the background in the Au layer is the highest followed by GaN and MgF$_2$. Therefore, the uncertainty of the resulting signal is larger in the Au layer especially for N and F which as light elements only give weak signals in EDX. For the 650\,$^{\circ}$C sample a decreasing signal of Mg and F extending roughly from the substrate into the GaN layer is visible, which is in qualitative agreement with recently reported SIMS data showing the intermixing without identifying the prevailing transport direction~\cite{Meyer2020a}. The EDX data show a small increase of Mg and F in the first 20-30~nm inside the GaN layer, but no accumulation of Ga or N within the substrate. The Mg and F signals just reach the 3$\sigma$ criterion as the usual detection limit for EDX. Hence, concentrations are estimated to be in the order of 0.5-1\%, which is well above their solubility limit in GaN indicating non-equilibrium conditions in this case \cite{Janotti2009,Liu2021,Neugebauer1995}. We may speculate that Mg and F incorporation mainly occurs at the very early stages of GaN growth and is related to partial MgF$_2$ decomposition on the nanoscale at high growth temperatures. In fact, cavity formation described above for the 650\,$^\circ$C sample can be viewed as advanced states of such decomposition. For the 525\,$^{\circ}$C sample no in-diffusion of Mg and F can be observed with respect to the EDX profiles. The interface between GaN and the Au capping layer deposited prior to TEM sample preparation is broader at the 650\,$^{\circ}$C sample compared to the 525\,$^{\circ}$C sample. It should be noted, that this difference originates from the larger surface roughness of the GaN layer grown at 650\,$^{\circ}$C.

\subsection{4D-STEM nanobeam electron diffraction data and their analysis}\label{sec033}

\begin{figure}
\centering
\includegraphics[width=.98\columnwidth]{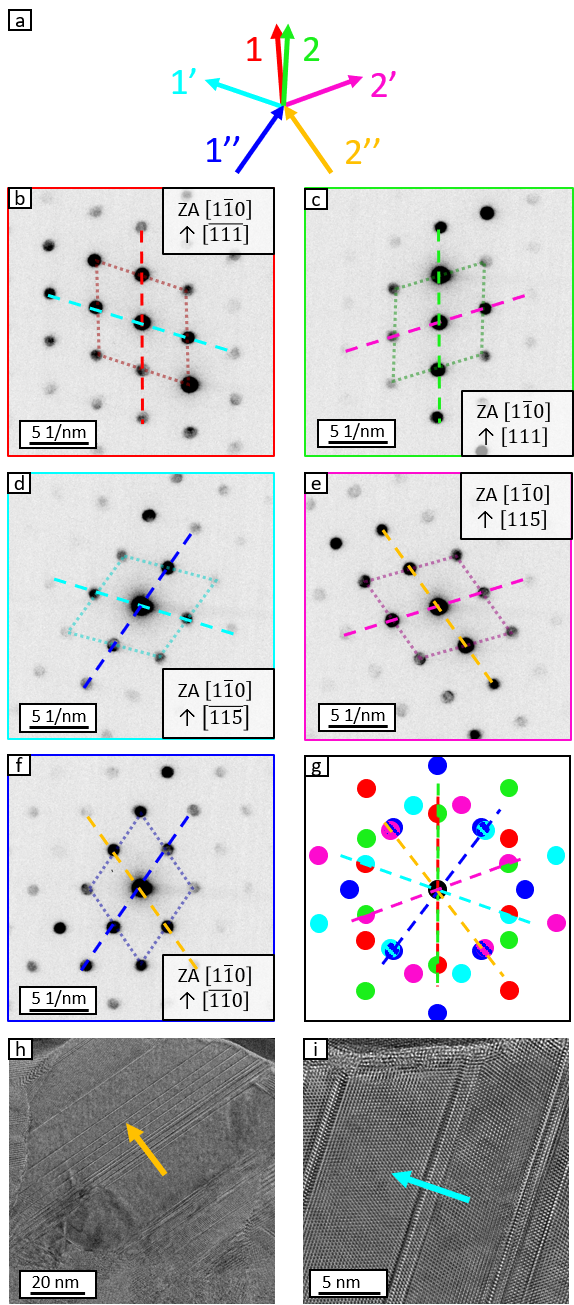}
\caption{NBED patterns (b)-(f) and HRTEM images (h)-(i) of the 525\,$^{\circ}$C sample with ZA [001] of MgF$_2$ substrate. The arrows of the scheme in (a) indicates directions of (111)-plane normals where twinning has been observed. To visualize the mirror axes of the twinning (g) shows an overlay of calculated diffraction patterns with the spots colored corresponding to the frames of the experimentally observed NBED patterns. The dashed rhombus in (b)- (f) are a guidance to see the orientation and the dashed lines are the normals of the (111)-planes where twinning has been observed.}
\label{figure3}
\end{figure}

In this section, we describe results on the spatial distribution of different c-GaN orientations. Since the observed orientations in the experimentally obtained 4D-STEM data are the same for both growth temperatures, the different orientations will be explained at the 525\,$^{\circ}$C sample (Sec.~\ref{sec0331}). The NBED patterns of the 650\,$^{\circ}$C sample are shown in the supplement as well as diffraction patterns of the substrate. To compare the distribution of the orientations and to analyze the grain size virtual dark-fields will be used (Sec.~\ref{sec0333}). The second part of Sec.~\ref{sec0333} is devoted to discuss the consistency of the 4D-STEM data for both zone axes [010] and [001] of the substrate as well as by reciprocal space mapping (RSM) as a supporting experiment (Sec.~\ref{sec034}).
%=======================================================================

\subsubsection{Twinning-related orientations of cubic GaN} \label{sec0331}

\begin{figure*}[htp]
\centering
\includegraphics[width=.96\textwidth]{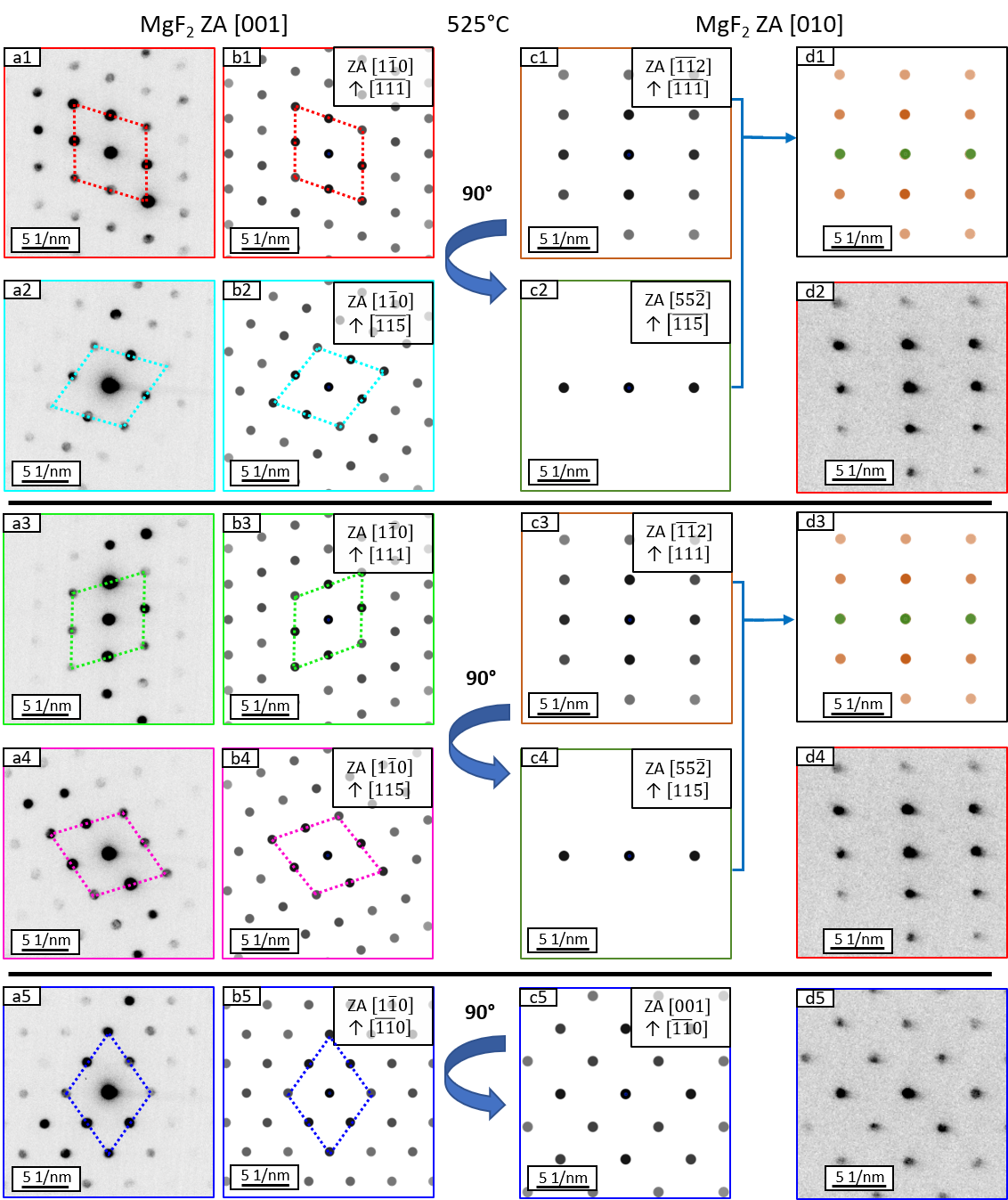}
\caption{Experimental NBED patterns of the 525\,$^{\circ}$C sample with their associated simulated pattern. (a1)-(a5) are along the [001] ZA of MgF$_2$ and (d2), (d4) and (d5) are along the [010] ZA of MgF$_2$. The dashed rhombuses are a guide to the eye to assist identification of the different orientations. Arrows indicate the change of the view direction from the [001] to the [010] zone axis which is a 90\,$^{\circ}$ rotation around the surface normal. (b1) and (b3) have the same pattern in the other direction as well as (b2) and (b4). The shown superposition of the diffraction patterns (c1) and (c2), (c3) and (c4) results from alternating twin lamellae of these orientations along the [010] MgF$_2$ direction on a nanometer length scale. The color code is individual for each substrate orientation.}
\label{figure4}
\end{figure*}

To identify the crystallographic orientations of the GaN layer, 4D-STEM is used which creates a stack of NBED patterns where each diffraction pattern corresponds to a pixel in real space. Most of the NBED patterns from the 525\,$^{\circ}$C sample along the zone axis [001] of MgF$_2$ show c-GaN along $\langle$110$\rangle$ zone axes with five different orientations, which are related by twinning. Representative NBED patterns of the five mainly observed c-GaN orientations of this sample are shown in Fig.~\ref{figure3}. All of the five different orientations (Fig.~\ref{figure3}(b)-(f)) share a common $\langle$110$\rangle$ zone axis of c-GaN, the dashed rhombuses are a guide to the eye for the different orientations. Along the \{111\} plane normals (dahed lines) twinning takes places. This results in five different surface normal orientations for the c-GaN along [111], [$\bar{1}\bar{1}\bar{1}$], [115], [$\bar{1}\bar{1}\bar{5}$] and [$\bar{1}\bar{1}0$]. In Fig.~\ref{figure3}(a) the arrows indicate the direction of the \{111\} plane normals of the different orientations. The numbering indicates the order of twinning and is not related to which orientation grows first. The orientations belonging to 1 (red) and 2 (green) are twins having a [$\bar{1}\bar{1}\bar{1}$] and a [111] surface normal, the corresponding NBED patterns are shown in Fig.~\ref{figure3}(b) and (c) respectively. 1' (cyan) and 2' (magenta) with [$\bar{1}\bar{1}\bar{5}$]/[115] surface normal are twins of 1 (red) and 2 (green), respectively, NBED patterns are given in Fig.~\ref{figure3}(d) and (e). The normals 1'' (blue) and 2'' (orange) belong to the same orientation with [$\bar{1}\bar{1}0$] surface normal as shown in Fig.~\ref{figure3}(f) and represent twins of 1' (cyan) and 2' (magenta), respectively, which identifies them as being 2$^{nd}$ order twins of 1 and 2, respectively. 

The twinning can also be observed in the HRTEM images Fig.~\ref{figure3}(h) and (i) where the arrows indicate the \{111\} plane normal, the other orientations are also observed in HRTEM and are not shown for brevity. It is important here to note that there is a small misfit between \{111\} planes for higher order twins of $\approx$ 3.68\,$^{\circ}$ which is visible in Fig.~\ref{figure3}(a) as the 1 (red) and 2 (green) arrow do not coincide in the overlaid calculated diffraction patterns in Fig.~\ref{figure3}(g) at the spots on the blue and orange dashed lines. It should be noted that the angular tilt can occur between any of the orientations since it is only determined by the initial twinning. A similar kind of misfit is known from fivefold twinning on \{111\} planes in fcc-metals, i.e. Zhu et al. \cite{Zhu2005}. This is in accordance with our experimental data as a rotation of a few degrees for NBED patterns of the same orientation can be detected at different positions of the 4D-data stack.

For c-GaN,  twinning in general is possible on all four $\{111\}$ planes. Experimentally, twinning on only two $\{111\}$ planes is directly observed, namely those who share the c-GaN~$[1\overline10]$ direction parallel to MgF$_2$~[001]. As shown in the supplement, contributions from the other $\{111\}$ planes cannot be directly observed in the given geometry as the diffraction spots overlap with the other orientations. Nevertheless,  an estimation can be made by vDF images as explained in Sec. \ref{sec0333}.

% %=======================================================================
\subsubsection{\texorpdfstring{Analysis of NBED data along MgF$_2$ [010] and [001] cross sections}{Analysis of NBED data along MgF2 [010] and [001] cross sections}}\label{sec0332}

\begin{figure*}[ht]
\centering
\includegraphics[width=.98\textwidth]{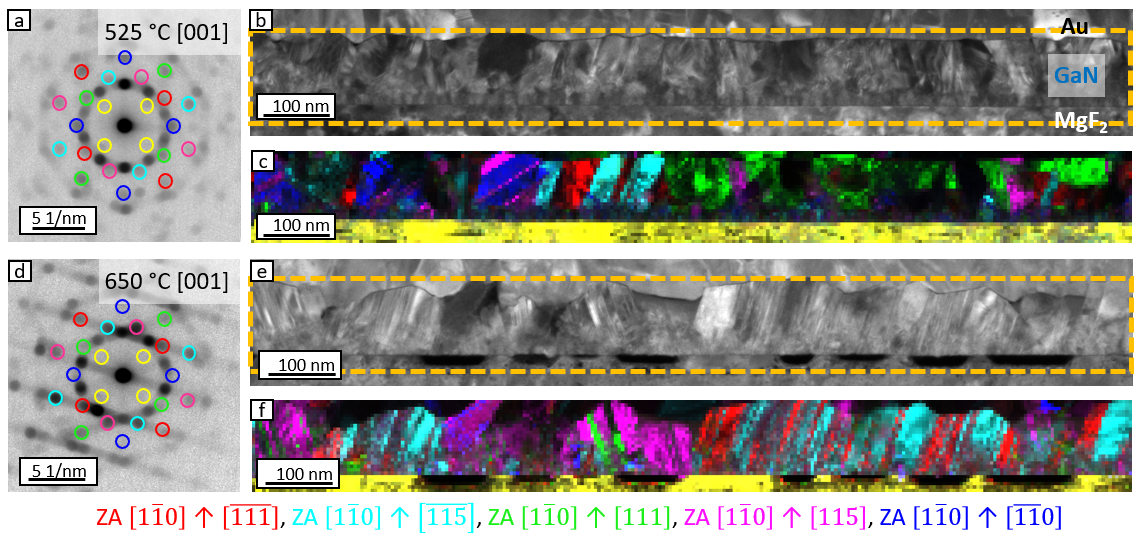}
\caption{4D-STEM results of lamellae with [001] MgF$_2$ ZA of both growth temperatures. The logarithm of the summed NBED patterns from the GaN layer are shown in (a) and (d) where the colored circles mark the masks which are used to create the vDF of each orientation. The color code is the same as in Fig.~\ref{figure4}, the yellow circles are the mask for the substrate. (b) and (e) show the ADF reference image with the orange frame marking the scan region. The superposition of the five vDFs for each orientation is given in (c) and (f) and represents a real space map for the different GaN orientations. The MgF$_2$ substrate is colored in yellow in the vDF.}
\label{figure5}
\end{figure*}

The 4D-STEM studies reported in this paper have been obtained in cross-section along the MgF$_2$ [001] and [010] zone axis. To show the relation of the different observed orientations for the two zone axes, the mainly observed NBED patterns and their corresponding simulated pattern are shown in Fig.~\ref{figure4}. For the purpose of NBED pattern simulations we use the simulation software JEMS~\cite{Stadelmann2003}. As the orientations are the same for both growth temperatures this is shown here for the 525\,$^{\circ}$C sample, the NBED patterns for the 650\,$^{\circ}$C sample are given in the supplement in Fig.~S2 in the same way. The essential difference between the 525\,$^{\circ}$C and 650\,$^{\circ}$C samples in terms of the diffraction patterns will be discussed in Sec.~\ref{sec0333}.

In Fig.~\ref{figure4} the starting point are the experimental NBED patterns of the GaN layer of the 525\,$^{\circ}$C sample with [001] zone axis of MgF$_2$, (a1)-(a5). As already shown in Fig.~\ref{figure3} all of the mainly observed NBED patterns have a $\langle$110$\rangle$ zone axis of c-GaN. In addition, the respective simulated pattern are shown in Fig.~\ref{figure4}(b1)-(b5). For those patterns, the prediction for the 90\,$^{\circ}$ rotated zone axis is made keeping the same growth direction, which is given in Fig.~\ref{figure4}(c1)-(c5). As the lamella with [010] MgF$_2$ ZA has a thickness of roughly 100\,nm and the twin lamellae with different orientations alternate along the view direction on a nanometer length scale, it is likely that a superposition of the simulated patterns Fig.~\ref{figure4}(c1) and (c2) will be observed which is represented by (d1). As the [$\bar{1}\bar{1}\bar{5}$] surface normal orientation (c2) has very few spots in the field of view which overlap with the spots in (c1), (c1) and (d1) look the same. The $\bar{1}\bar{1}\bar{5}$ spot is observed but due to rotating and cutting not visible in the presented frame, it is shown in the supplement in Fig.~S3. In Fig.~\ref{figure4}(d2), (d4) and (d5) the experimental NBED patterns of the lamella with [010] MgF$_2$ ZA are shown, which are both in agreement with the predicted simulated patterns in (d1), (d3) and (c5). For the lamella with the [010] MgF$_2$ ZA, these two patterns in Fig.~\ref{figure4}(d2)/(d4) and (c5), are the mainly observed patterns. It is important here to note that the experimental pattern in Fig.~\ref{figure4}(d2)/(d4) can easily be misinterpreted as h-GaN with a [1$\bar{1}$00] zone axis if there would not be the additional information from an other zone axis as the patterns have the same shape and the difference in the spacing of the first order spots is less than 0.05\,1/nm. Assigning the patterns in Fig.~\ref{figure4} (a1)-(a5) to c-GaN is straight forward as there is no possible h-GaN pattern being similar to the observed patterns.
%===========================================================
\subsubsection{Real space distribution of c-GaN orientations from virtual dark field imaging}\label{sec0333}

\begin{figure*}[ht]
\centering
\includegraphics[width=.98\textwidth]{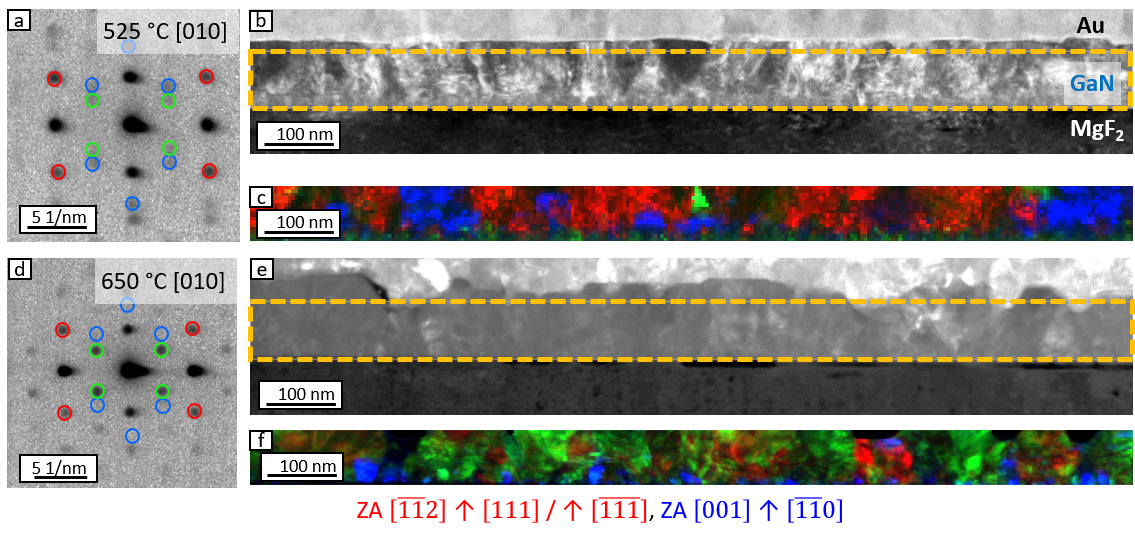}
\caption{4D-STEM results of lamellae with [010] MgF$_2$ ZA of both growth temperatures. The logarithm of the summed NBED patterns from the GaN layer are shown in (a) and (d) where the colored circles mark the masks which are used to create the vDF of each orientation. The color code is the same as in Fig.~\ref{figure4} where green marks the additional observed spots. (b) and (e) show the ADF reference image with the orange frame marking the scan region. The superposition of the three vDFs for each set of masks is given in (c) and (f) and represents a real space map for the different GaN orientations.}
\label{figure6}
\end{figure*}

Using the stack of NBED patterns of a 4D-STEM experiment, virtual dark field (vDF) images can be generated by masking certain spots and display the summed intensity inside the mask as a function of electron beam position in real space. The knowledge of the primary orientations of the GaN layers is necessary to create a mask for each orientation exclusively containing spots for this orientation, Fig.~\ref{figure3}(g) helps to find such spots. These masks are applied to the stack of NBED patterns resulting in a vDF for each orientation. Finally a superposition of the virtual dark fields is created where each vDF has an individual color. The combined vDF shows then a map of the different orientations in the GaN layer. In Fig.~\ref{figure5} the results of this procedure are shown for the 525\,$^{\circ}$C and the 650\,$^{\circ}$C samples with the [001] zones axis of MgF$_2$. Fig.~\ref{figure5}(a) and (d) show the logarithm of the summed NBED patterns of the GaN layer where the colored circles indicate the five masks for each orientation where the color code is the same as in Fig.~\ref{figure3} and Fig.~\ref{figure4} and the MgF$_2$ substrate is colored in yellow. The ADF reference image is given in Fig.\ref{figure5}(b) and (e) and with the dashed orange frame the scan area is marked. The resulting real space map of the GaN orientations, the superposition of the virtual dark fields, is given in Fig.~\ref{figure5}(c) and (f).

The maps of the GaN orientations in Fig.~\ref{figure5}(c) and (f) clearly show the twinning relationship between the different orientations of the c-GaN. The red and cyan stripes in the right half of Fig.~\ref{figure5}(f) may serve as a pronounced example. At the interface between MgF$_2$ and the GaN layer no preferred orientation is observed and the first 10-15\,nm do not show a clear orientation. This behaviour is more pronounced at the 525\,$^{\circ}$C sample. The GaN orientation, which has the [$\bar{1}\bar{1}0$] surface normal (colored blue) is only found in a few patterns close to the interface for the 650\,$^{\circ}$C sample in contrast to the 525\,$^{\circ}$C sample, where larger areas of this orientation are observed in the whole GaN film. Comparing both growth temperatures, a higher density of twins - equivalent to thinner twin lamellae -  can be observed at the 650\,$^{\circ}$C sample. The dark regions at the top of the GaN layer belong to the Au layer and should be ignored. The darker parts inside the GaN layer are related to the fact that there is not always one clear orientation visible and in addition the masking takes certain spots with high intensity not into account if they are not exclusive for one orientation. As the intensity of the spots vary due to orientation fluctuations and defects, the intensity is not homogeneous in the vDF. Based on a vDF created by an annular mask around the $\{111\}$ spots from the 525\,$^{\circ}$C sample with [001] MgF$_2$ ZA 11\,\% of the GaN layer can be estimated as dark regions which cannot be identified. For the 650 \,$^{\circ}$C sample with [001] MgF$_2$ ZA there are less dark regions, approximately 2\,\%. One possible origin for the dark region are twins from the remaining two $\{111\}$ planes that are not directly observed.

In Fig.~\ref{figure6} the 4D-STEM results of both growth temperatures with the MgF$_2$ zone axis of [010] are shown in the same way as in Fig.~\ref{figure5}. As already discussed at Fig.~\ref{figure4} for this cross section orientation mainly two different diffraction patterns can be observed as four of the five different orientations have in principle the same diffraction pattern when viewed along this direction. The c-GaN orientation having the [$\bar{1}\bar{1}0$] surface normal is the only orientation with a distinct diffraction pattern in this view direction. In Fig.~\ref{figure6} the masks and the vDFs of this (110) c-GaN are colored in blue. The c-GaN with the [$11\bar{2}$] and [$\bar{1}\bar{1}2$] zone axes are colored in red which includes both orientations with (111) surface normal (i.e. the green and red orientations in Fig.\,\ref{figure5} ). For the 650\,$^{\circ}$C sample the (110) c-GaN can only be found in small grains close to the interface to the MgF$_2$ where it can be found for the 525\,$^{\circ}$C sample in the whole GaN film. This is in good agreement with the other view direction. Instead, additional spots are observed in the 650\,$^{\circ}$C sample, which are masked in green in Fig.~\ref{figure6}(a) and (d). These patterns are compatible with h-GaN along the  [0001] zone axis, an identification being, however, incompatible with diffraction data obtained along MgF$_2$ [100] as described above.

\begin{figure}
\centering
\includegraphics[width=.97\columnwidth]{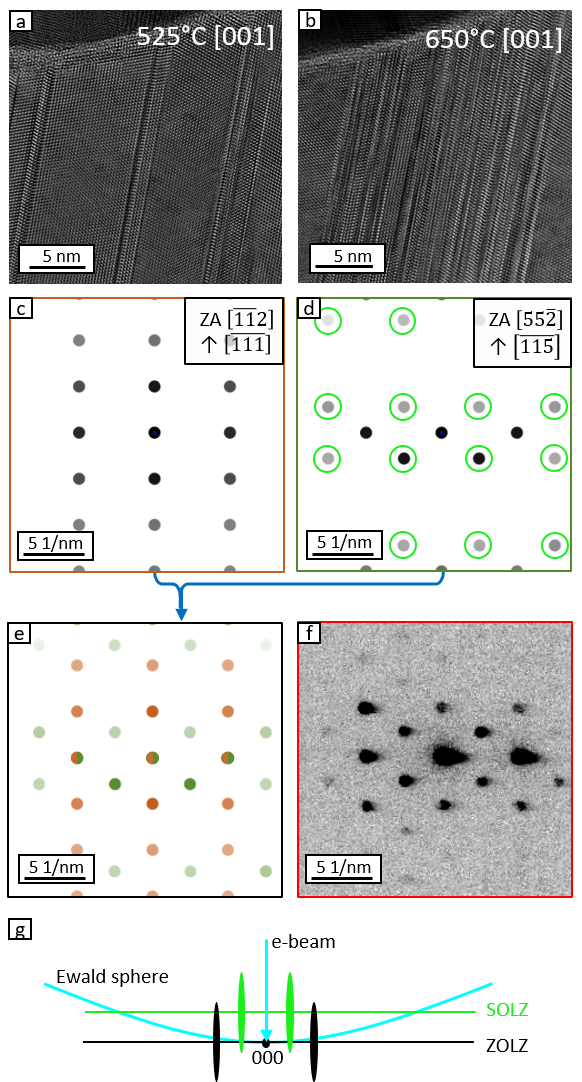}
\caption{The HRTEM images in (a) and (b) show the difference of the density and distance of the twin boundaries based on the different growth temperature. (c) and (d) are simulated patterns which include the SOLZ, the additional spots are marked with green circles. The superposition of (c) and (d) is shown in (e) which is in agreement with experimental NBED pattern of the 650,$^{\circ}$C sample with [010] zone axis of MgF$_2$ in (f). (g) illustrates schematically the elongation of the relrods due to the small distance of the twin boundaries in (b) which results to an intersection of the Ewald sphere of the spots from the SOLZ.}
\label{figure7}
\end{figure}

For a consistent interpretation of these diffraction spots, we have to consider the situation for grains with [$\bar{1}\bar{1}\bar{5}$] along the growth direction in some detail. As mentioned above, these grains are viewed along a [55$\bar{2}$] zone axis in cross sections taken along the MgF$_2$ [010] direction. For bulk samples, diffraction spots of $2\bar2 0$ and 115 (and their linear combinations) define the zero order Laue zone (ZOLZ). As mentioned in Fig.\,\ref{figure4} and related text, the spots overlap with spots from the [$11\bar{2}$] zone axis patterns of grains with (111) surface normal as shown again by the simulated diffraction patterns in Fig.\,\ref{figure7} (c) and (d). Since the [55$\bar{2}$] zone axis is a high-indexed direction, the related second order Laue zone (SOLZ) has quite a small distance to the ZOLZ, i.e. only  D$_{SOLZ}\approx$ 0.6\,nm$^{-1}$. The simulated diffraction pattern in Fig.\,\ref{figure7}(d) also contains spots from the SOLZ (marked by green circles), which can be indexed as $1\bar1\bar1$, $\bar1 1\bar1$ and their linear combinations with reciprocal lattice points from ZOLZ. These spots exactly match the experimentally observed spots, as revealed by comparing Figs.\,\ref{figure7} (e) and (f). It remains to be explained why spots from SOLZ appear in experimental diffraction patterns: a straightforward explanation is a shape effect arising from the atomically thin twin lamellae observed in particular in the 650\,$^\circ$C sample, where in fact the spots are much stronger and more frequent compared to the 525\,$^\circ$C sample. Due to the reduced extension of these nanotwins, their related reciprocal lattice spots are extended reciprocal lattice rods ('relrods' \cite{WilliamsCarter}) rather than points leading to diffraction intensity if the twin lamellae are sufficiently thin (Fig.\,\ref{figure7}(g)). We can roughly estimate the relrod's extension as (n$\cdot d_{111})^{-1}$, where n$\cdot d_{111}$ is the twin lamella thickness expressed as multiples of the (111) lattice plane spacing. Comparing this to D$_{SOLZ}$ gives an upper limit for n as n$_{max}\approx$ (d$_{111}$D$_{SOLZ}$)$^{-1}\approx$ 6, in other words: nanotwins with a thickness smaller than six (111) lattice planes are expected to produce intensity in a [55$\bar2$] diffraction pattern. Inspection of the HRTEM images in Fig.\,\ref{figure7}(a) and (b) provides evidence for a high density of sufficiently thin nanotwins in the 650\,$^{\circ}$C sample and a smaller density in the 525\,$^{\circ}$C sample, which, in fact, only occasionally shows these spots.

As a preliminary summary, NBED data obtained along two different cross-section directions (along MgF$_2$ [010] and [001]) and their analysis presented in Sec.\,\ref{sec033} consistently show growth of c-GaN on MgF$_2$ [100] substrates without noticeable contribution from h-GaN. The observed different c-GaN orientations are related via twin transformations and share a common c-GaN $\langle$110$\rangle$ direction parallel to MgF$_2$ [001].

\subsection{XRD reciprocal space mapping}\label{sec034} 
\begin{figure*}
\centering
\includegraphics[width=.97\textwidth]{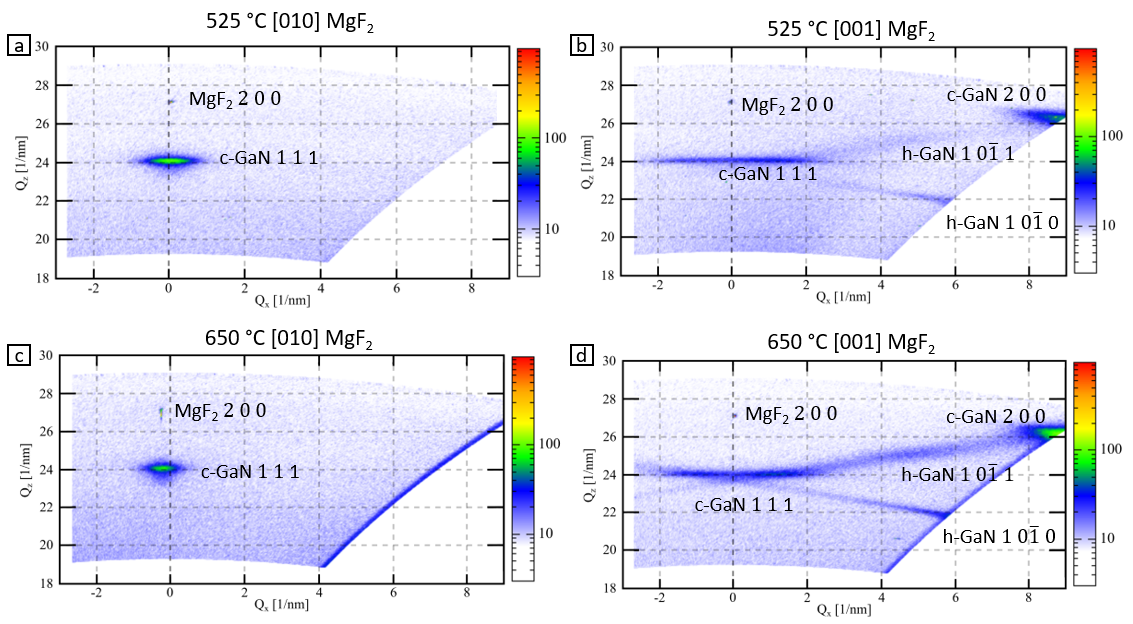}
\caption{Symmetric RSMs of both samples at $\phi$ values of 0\,$^{\circ}$ (along [010] MgF$_2$) and 90\,$^{\circ}$ (along [001] MgF$_2$): The 525\,$^{\circ}$C sample has only one film related RLP, the 111 RLP of c-GaN (111) in a), while additional RLPs are detected after a rotation of 90\,$^{\circ}$. These orientations are highly tilted to the substrate. RSMs of the 650\,$^{\circ}$C sample in (c) and (d) have no differences compared to 525\,$^{\circ}$C.}
\label{figure8}
\end{figure*}

Besides the valuable information on the microstructural properties of the GaN thin films on MgF$_2$, our NBED data partly disagree with XRD analyses reported by Meyer et al.~\cite{Meyer2020a,Meyer2020a-corr} as the NBED data indicates predominant growth of c-GaN instead of h-GaN. It is difficult to distinguish between the two phases in XRD as for example the 111 c-GaN and the 0002 h-GaN reflections nearly have the same $2\theta$ angle \cite{Frentrup2017}. In order to check this apparent discrepancy, symmetric reciprocal space maps (RSM) have been recorded (see Fig.~\ref{figure8}) at $\phi$-values of 0\,$^{\circ}$ and 90\,$^{\circ}$.

In Fig.~\ref{figure8}(a), the symmetric RSM of the 525\,$^{\circ}$C sample is presented. The RSM has two reciprocal lattice points (RLP), which are the 200 RLP of MgF$_2$ and the 111 RLP of c-GaN. Two 360°-$\phi$-scans were additionally performed for two asymmetric RLPs to differentiate between c-GaN and h-GaN. These $\phi$-scans (not shown) indicate a formation of c-GaN, so this sample orientation matches the results of the STEM analyses for the lamella with [010] MgF$_2$ zone axis. c-GaN has no tilt to the substrate. After a rotation of the sample by 90\,$^{\circ}$ in $\phi$-direction, several tilted RLPs are detected. This is shown in Fig.~\ref{figure8}(b). At Q$_x$=0, the 111 RLP of c-GaN seems now to be split into two RLPs which can be interpreted as an indirect evidence for the higher order twinning. As explained in Sec. \ref{sec0331} second order twinning leads to a tilt of $\approx 3.68$\,$^{\circ}$. The observed splitting angle is in the same order indicating that the origin of the splitting is based on this misfit. An in-plane 360\,$^{\circ}$-$\phi$-scan of c-GaN (111) (not shown) illustrates a 6-fold symmetry proving the formation of twins.
The additional tilted RLPs can be associated to c-GaN (100), h-GaN (10$\overline{1}$1) and possibly h-GaN (10$\overline{1}$0). The line-shaped intensity between the 111 and 200 RLP indicates a high density of extended defects. As expected the RSM of the 90$^{\circ}$ rotated sample matches the NBED patterns of the lamellae with [001] MgF$_2$ zone axis. Due to the high tilt of these three orientations, these could not be found in out-of-plane scans in $\theta$-$2\theta$ geometry in~\cite{Meyer2020a,Meyer2020a-corr}.
The symmetric RSMs of the 650\,$^{\circ}$C sample are shown in Fig.~\ref{figure8}(c) and (d). There are the same two RLPs at $\phi$=0\,$^{\circ}$ and five RLPs at $\phi$=90\,$^{\circ}$, indicating the formation of the same orientations at both growth temperatures. The split of the 111 RLP of c-GaN (111) is now clearly visible. The 220 and 115 c-GaN RLP could not be measured in the range of this RSM. Both RLPs could be found in further RSMs (not shown) without a change after a rotation of the sample.

Both the RSMs and the NBED patterns show the reciprocal space. By comparing them to each other it is important to note that the scaling differs by a factor of $2\pi$ due to different conventions. Due to the geometry of the experimental setup, RSM mainly captures reflections close to the surface normal, i.e. the field of view is limited compared to NBED. Therefore, the substrate orientation in the in-plane direction is not a direct result of the RSMs. Nevertheless, the orientation of the sample in the RSM and the TEM lamella orientation can be chosen the same leading to a perfect match of the observed reflections from both measurements. Regarding the 10$\overline{1}$1 and the 10$\overline{1}$0 h-GaN reflections in the RSM no spots but only streaks are observed between for example the 111 and 200 spots in the NBED patterns.

\subsection{Epitaxial relationships}\label{sec035}
As shown in Sec.~\ref{sec033}, virtually \textit{all} observed electron diffraction patterns can be consistently described in terms of c-GaN (lower limits are 89\%  and 98\% for growth at 525$^\circ$C and $650^\circ$C, respectively). Different orientations share the c-GaN~$[1\overline10]$ direction, which is observed parallel to MgF$_2$~[001]. These orientations are related to each other via (successive) twinning on those $\{111\}$ planes containing $[1\overline10]$, i.e. $(111)$ and $(11\overline1)$ (an extensive treatment is provided in the Supplemental Material). This is, however, a rather formal sight on the results, which does not allow for drawing conclusions related to the predominant or initial epitaxial relationships between (100)~MgF$_2$ and c-GaN since processes \textit{during} growth cannot be deduced from the final result albeit for special cases.

\begin{table*}[ht]
\caption{Epitaxial relations observed in this work for c-GaN growth on (100) MgF$_2$. The interplanar mismatch $\delta$ has been calculated via $\delta = 1-d/D, $ where $d$ and $D$ are lattice plane spacings in GaN and MgF$_2$, respectively.   The common growth directions in MgF$_2$ and GaN are  denoted as UVW  and uvw,  respectively. $^\dag$ Note, that there is small misfit ($\approx 0.1\%$) comparing 5$\times$($\overline1\overline12$)-GaN and 4$\times$(020)-MgF$_2$.} \label{table02}
\begin{center}
\begin{tabular}{|l|ccl|ccl|r|}
\hline
\multirow{2}{*}{phase / epitaxy} & \multicolumn{3}{c|}{ MgF$_2$} & %
    \multicolumn{3}{c|}{ GaN } & \multirow{2}{*}{$\delta$ [\%]}\\
\cline{2-7}
% & \multicolumn{2}{c|}{Value} & \multicolumn{2}{c|}{Value} & \\
%\cline{2-5}
& UVW & HKL & D [nm] & uvw & hkl & d [nm] & \\
\hline
\multirow{2}{*}{c-GaN / 2/2 epitaxy} & \multirow{2}{*}{100} & 020& 0.2310 & \multirow{2}{*}{110}  & 002 & 0.2260 & 2.2\\
%\cline{2-6}
%&  & 020 & 0.231 & & 002  & 0.226 & 2.2\\
%\cline{2-6}
& & 002 & 0.1525 & & 2$\Bar{2}$0  & 0.1598 & -4.8 \\
\hline
\multirow{2}{*}{c-GaN / 2/3 epitaxy } & \multirow{2}{*}{100} & 020 & 0.2310 & \multirow{2}{*}{111}  &  $\overline1\overline12$ & 0.1846 & 20.0$^\dag$\\
& & 002 & 0.1525 & & $2\overline20$  & 0.1598 & -4.8 \\
\hline
\multirow{2}{*}{c-GaN / 2/$\overline3$ epitaxy } & \multirow{2}{*}{100} & 020 & 0.2310 & \multirow{2}{*}{$\overline1\overline1\overline1$}  &  $\overline1\overline12$ & 0.1846 & 20.0$^\dag$\\
& & 002 & 0.1525 & & $2\overline20$  & 0.1598 & -4.8 \\
\hline
\end{tabular}
\end{center}
\end{table*}

\begin{figure}
\centering
\includegraphics[width=\columnwidth]{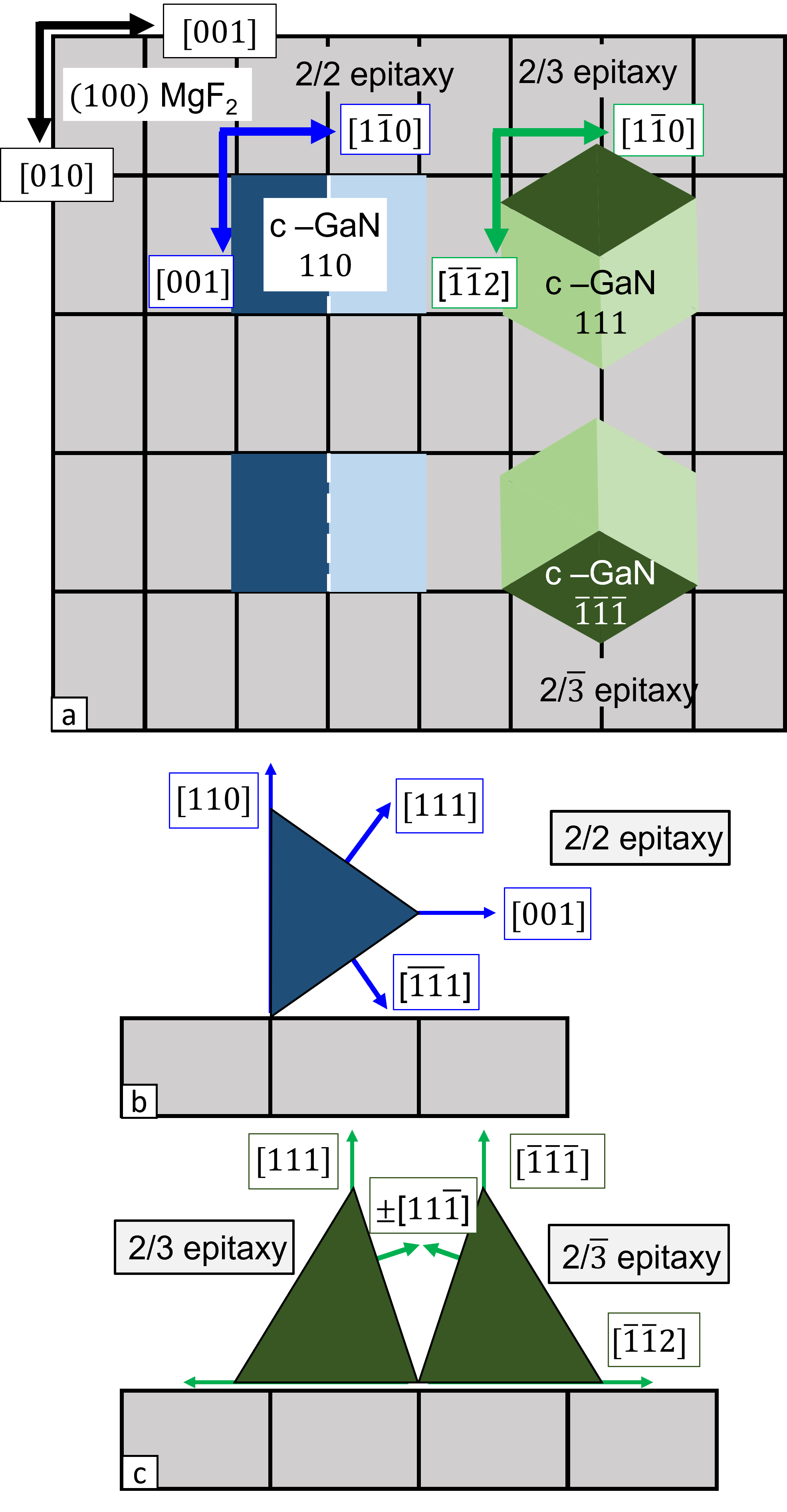}
\caption{Schemes showing epitaxial relationships and crystallographically given facts important for description of twinning. (a) top view along (100)~MgF$_2$, (b) and (c): cross-section views along (001)~MgF$_2$ $\parallel$ c-GaN~$(1\overline10)$ in all cases for the 2/2-epitaxy and 2/3- (or 2/$\overline3$)-~epitaxy), respectively, representing the different orientations in terms of \textit{Thompson tetrahedra}.  As outlined in the text, the notation highlights the symmetry of the substrate (2-fold) and the epilayer, which is 2-fold for $(1\overline10)$~c-GaN (blue in (a)) and 3-fold for (111)~c-GaN (green in (a)). As a consequence, the epitaxially equivalent variants produced by 180$^\circ$ rotation about the substrate normal are either fully equivalent (for the 2/2-epitaxy) or are related by a twin operation as for the 2/3- and 2/$\overline3$-epitaxy.}
\label{figure9}
\end{figure}

 In this section, epitaxial relationships consistent with the observed extensive twinning will be discussed in more detail. For classification, reference to the symmetry of the substrate (2-fold) and of the c-GaN layer in different orientations (2-fold or 3-fold) will be considered. 
Experimentally, cross-sections along MgF$_2$ [010] and [001] have been analyzed revealing two main orientational relationships: GaN~(110) $\parallel$ MgF$_2$~(100) with GaN~[001]$\parallel$MgF$_2$~[010] and GaN~[$1\overline10$]$\parallel$MgF$_2$~[001] (in short: $[1\overline10](110)$~c-GaN $\parallel$ [001](100)~MgF$_2$)
and GaN~(111)$\parallel$MgF$_2$~(100) with GaN~$[\overline1\overline12]$ $\parallel$ MgF$_2$~[010] and GaN~[$1\overline10$]$\parallel$MgF$_2$~[001] (in short: $[1\overline10](111)$c-GaN $\parallel$ [001](100)MgF$_2$). In the former case, the 2-fold symmetry of the substrate is preserved in the layer; this mode will be referred to as '2/2-epitaxy'. In the latter case, however, a layer with 3-fold symmetry grows indicating that two variants will occur, which are equivalent from the point-of-view of epitaxial growth, but are different when taking into account out-of-plane directions. These two variants will be referred to as '2/3-epitaxy' and 2/$\overline3$-epitaxy as will become clear when referring to Fig.~\ref{figure9} showing top views (along MgF$_2$~[100]) in (a) and side views (along MgF$_2$~[001] and along c-GaN~$[1\overline10]$) in (b) and (c), where \textit{Thompson tetrahedra} are used to show the $\{111\}$ planes relevant for twinning. The 2/3- and 2/$\overline3$- epitaxy in Fig.~\ref{figure9}(c) lead to an angular tilt of $220$ diffraction spots from secondary twins by $\approx\pm3.68^\circ$ from the substrate normal (compare Fig.~S5 in Supplemental Material). Similarly, secondary twins in case of the 2/2-epitaxy reveal $(111)$ and $(11\overline1)$ diffraction spots also tilted by $\approx\pm 3.68^\circ$ from the substrate normal (compare Fig.S4 in Supplemental Material), which explains the (111) spot splitting in RSM mentioned in Sec.~\ref{sec034}. 

It is tempting to propose the 2/2-epitaxy as being the 'underlying' growth mode for two reasons, i.e. (i) the relatively small misfit compared to the 2/3- and 2/$\overline3$ -~epitaxy (see Table~\ref{table02}) and (ii) the observation of the related crystal orientation predominantly located close to the substrate for the 650$^\circ$C sample (Fig.~\ref{figure6}). We want to stress here, however, that this is more a speculation rather than a conclusion, since according to our 4D-STEM data also the 220 diffraction spots show angular tilts possibly originating from the 2/3- and 2/$\overline3$-epitaxy as outlined above. In order to clarify this point additional growth experiments and extensive X-ray reciprocal space mapping as well as NBED studies are required, which is beyond the scope of this paper.

\section{Summary and Conclusion}
Successful epitaxial growth of c-GaN on MgF$_2$ offers the intriguing perspective of solarblind photodetectors directly deposited on a first class window material. For growth on MgF$_2$ (100) substrates, our combined NBED and XRD RSM data congruently reveal c-GaN growth with growth directions of $\langle 110\rangle$, $\langle 111\rangle$, and $\langle 115  \rangle$ as unambiguously shown by 4D-STEM NBED along two different cross-section directions (along MgF$_2$ [010] and [001]) (see Secs.\,\ref{sec033} and \ref{sec034}).The observed different c-GaN orientations are related via twin transformations and share a \textit{common} c-GaN $\langle$110$\rangle$ direction parallel to MgF$_2$ [001]. From lattice plane spacings it follows that there is an interplanar mismatch along this \textit{common} direction of -4.8\%, i.e. tensile strain.

Growth of c-GaN has already been performed on different substrate materials. For example, growth on (001)-GaAs has been investigated by by Qu et al.~\cite{Qu2001} showing c-GaN with (001) surface normal and twinning on \{111\} planes. In a similar way, c-GaN growth occurs on (001)-3C SiC \cite{Lee2019}. For this situation, the formation of twins has been related to relaxation of strain resulting from the coalescence of islands initially nucleated on (001) cubic substrates \cite{Trampert1997}. These studies nicely show the tendency of c-GaN to form twins as an energetically favourable mechanism of strain relaxation. In fact, from the point of view of the stacking sequence of (111) planes, twin boundary can be viewed as a monolayer of h-GaN. This might be related to the intensity at h-GaN $10\bar{1}1$ and $10\bar{1}0$ reciprocal lattice point observed in our reciprocal space maps along the [001] MgF$_2$ direction (\ref{figure8}). This view is corroborated by the higher intensity around these spots for the 650\,$^{\circ}$C sample compared to the 525\,$^{\circ}$C sample as the twin density is much higher in the former case. We mention here, that besides at the twin boundaries a relevant amount of h-GaN is not observed in our system as was previously observed for the (001)-GaAs/GaN system close to c-GaN nanotwins \cite{Qu2001}.

We note here, that similar growth of (110) c-GaN has been reported for GaN epitaxy on (001) rutile TiO$_2$ by Araki et al. \cite{Araki2002}. Interestingly, rutile TiO$_2$ has the same structure and similar lattice parameters as MgF$_2$ (a=0.4593~nm and c=0.2959~nm for TiO$_2$ and 0.462~nm and 0.305~nm for MgF$_2$). Araki et al., however, studied c-GaN on the fourfold (001) surface in contrast to the twofold (100) MgF$_2$ surface studied in this work, so that a direct comparison is currently not possible. In fact, the twofold symmetry of (100) MgF$_2$ substrates is quite unique and delivers an interesting piece of information: As pointed out in Sec.\,\ref{sec033}, two out of the three c-GaN orientations parallel to MgF$_2$ [100] occur in two variants, i.e. the $\langle 111\rangle$ orientation (labeled 1 and 2 ) and the $\langle 115\rangle$ orientation (labeled 1' and 2' in Fig.\,\ref{figure3}), which are related by a 180\,$^\circ$ rotation about the substrate normal thus reflecting the twofold symmetry of the substrate. For the $\langle110\rangle$ orientation such a 180\,$^\circ$ rotation is a symmetry operation for the cubic $\langle110\rangle$ direction. This has led to introduce the terms '2/2-epitaxy and '2/3- (or 2/$\overline3$)-epitaxy for $[1\overline10](110)$~c-GaN $\parallel$ [001](100)~MgF$_2$ and $[1\overline10](111)$c-GaN $\parallel$ [001](100)MgF$_2$ orientational relationships, respectively.
Hence, the whole twinning microstructure is in line with the twofold symmetry of the underlying substrate. We note here, that \textit{formally} the five orientations have identical geometry as observed in \textit{fivefold twinned} nanostructures of fcc materials\,\cite{Zhu2005}. Our 4D-STEM virtual dark field images (Fig.\,\ref{figure5}), however, show that these five twin orientations are observed in different domains rather than in the same region.

Notwithstanding the fact that our experiments have not been designed to study the nucleation of c-GaN on MgF$_2$ (100), let us consider the observation that [110] oriented grains exclusively occur at the layer-substrate interface for growth at 650\,$^\circ$C (Fig.\,\ref{figure6}). Assuming these grains to be reminiscent of the nucleation stage, let us consider the misfit with the substrate, which is  -4.8\% along [$1\bar 10$] and +2.2\% along [001]. Hence, the resulting stress is tensile along [$1\bar 10$] \textit{common to all} observed orientations and compressive perpendicular to it. Strain relaxation by forming first and second order twins keeping the common [$1\bar 10$] produces the observed GaN microstructure and possibly selects the two  out of the six possible twin systems. The important question whether this scenario is realized in this system has to remain open at this stage of research.

Finally, let us briefly refer to the striking observation of cavity formation inside MgF$_2$ substrates directly beneath the GaN for PAMBE growth at 650\,$^{\circ}$C. Their typical extension is 100~nm in lateral direction and 10~nm along the substrate normal and their existence witnesses the decreased stability of MgF$_2$ at this temperature leading to surface roughening and eventually cavity formation. The fact that GaN layers do not protrude into cavities indicates that the latter form during growth rather than prior to GaN deposition as a result of thermal decomposition of MgF$_2$. It remains open at this stage, whether cavity formation is related to Mg and F incorporation as detected by EDX and in agreement with previously published SIMS data \cite{Meyer2020a}. Although no microscopic model can be offered at this stage of research, both observations might be related to the observed high density of twinning as an increasing Mg concentration leads to increasing twinning in GaN nanowires as it was found by Aribol et al. \cite{Arbiol2009}.

In conclusion, we have presented an extensive scanning transmission electron microscopy study of the thin film microstructure for growth at 525\,$^{\circ}$C and 650\,$^{\circ}$C on (100) MgF$_2$ systematically supported by X-ray diffraction reciprocal space maps. The exclusively observed c-GaN occurs in five different orientations relate by twin transformations in line with the twofold symmetry of the substrate. For envisaged application of this system as solar blind photodetectors further growth optimisation is required with special focus on nanotwin formation and twin boundary related electronic states\,\cite{Bandic1997,Chu2009}.

\section*{CRediT authorship contribution statement}

\textbf{Tobias Niemeyer:} Conceptualization, FIB pre\-paration, TEM investigation, writing - original draft. \textbf{Kevin Meyer:} Growth of GaN layer, HRXRD investigation, writing - original draft. \textbf{Christoph Flathmann:} FIB preparation, TEM investigation. \textbf{Tobias Meyer:} TEM investigation, programming script for NBED acquisition and analysis. \textbf{Daniel M. Schaadt:} Supervision, writing - review \& editing. \textbf{Michael Seibt:} Supervision, writing - review \& editing.

\section*{Declaration of competing interest}

The authors declare that they have no known competing financial interests or personal relationships that could have appeared to influence the work reported in this paper.

\section*{Acknowledgments}

This work was partially funded by the Deutsche Forschungsgemeinschaft (DFG, German Research Foundation) under project no. 429413061. The use of equipment of the “Collaborative Laboratory and User Facility for Electron Microscopy” (CLUE, G\"ottingen) is gratefully acknowledged. The used XRD was financially supported by the Deutsche Forschungsgemeinschaft (DFG) in the framework of the INST 189/189-1 FUGG and is gratefully acknowledged.

\end{document}

\title{Microstructural analysis of GaN films grown on (1 0 0) MgF\texorpdfstring{$_{2}$}{2} substrate by 4D nanobeam diffraction and energy-dispersive X-ray spectrometry}

\author{Tobias Niemeyer}
\affiliation{4th Institute of Physics \textendash{} Solids and Nanostructures, University of Goettingen, Friedrich-Hund-Platz 1, 37077 G\"ottingen, Germany}
%\ead{tobias.westphal@uni-goettingen.de}
\author{Kevin Meyer}
\affiliation{Institute of Energy Research and Physical Technologies IEPT, Clausthal University of Technology, Leibnizstrasse 4, 38678 Clausthal-Zellerfeld, Germany}
\author{Christoph Flathmann}
\affiliation{4th Institute of Physics \textendash{} Solids and Nanostructures, University of Goettingen, Friedrich-Hund-Platz 1, 37077 G\"ottingen, Germany}
\author{Tobias Meyer}
\affiliation{Institute of Materials Physics, University of Goettingen, Friedrich-Hund-Platz 1, 37077 G\"ottingen, Germany}
\author{Daniel M. Schaadt}
\affiliation{Institute of Energy Research and Physical Technologies IEPT, Clausthal University of Technology, Leibnizstrasse 4, 38678 Clausthal-Zellerfeld, Germany}
\author{Michael Seibt}%\texorpdfstring{\corref{cor1}}{*}}
\email{michael.seibt@uni-goettingen.de}
\affiliation{4th Institute of Physics \textendash{} Solids and Nanostructures, University of Goettingen, Friedrich-Hund-Platz 1, 37077 G\"ottingen, Germany}

\maketitle

\section{Additional NBED patterns}\label{sec:SI1}

\begin{figure}[htp]
\centering
\includegraphics[width=.97\textwidth]{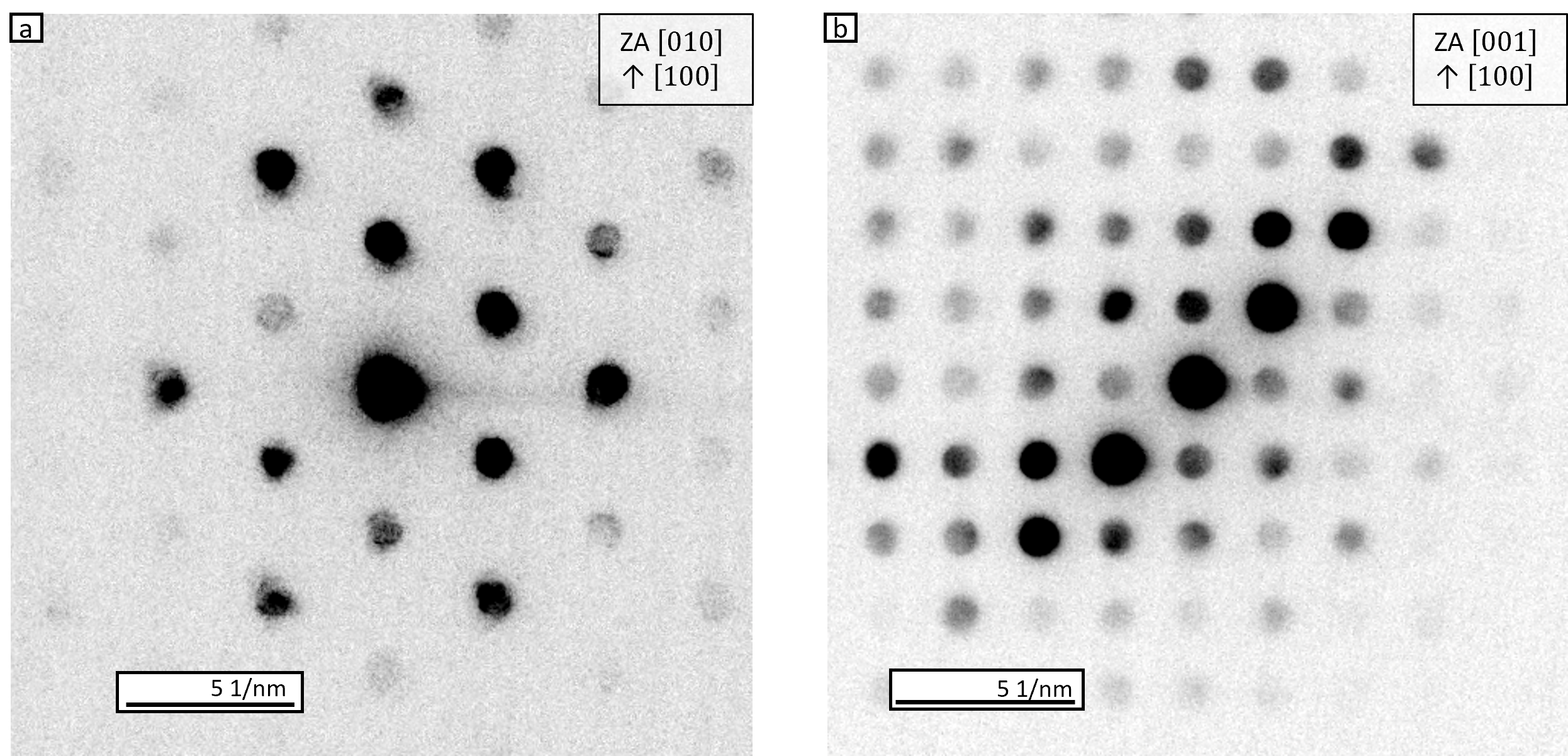}
\caption{NBED patterns of the MgF$_2$ substrate. (a) from the cross-section lamella prepared perpendicular to the [010] MgF$_2$ orientation and (b) form the lamella perpendicular to the [001] orientation of MgF$_2$. Both are taken from the 650\,$^{\circ}$C sample.}
\label{figureS1}
\end{figure}

As an addition to the presented NBED patterns in the main text in Fig.~\ref{figureS1} representative patterns of the MgF$_2$ substrate for both orientations are shown. These patterns confirm the orientation of the substrate as (100) MgF$_2$.

\begin{figure}[htp]
\centering
\includegraphics[width=.97\textwidth]{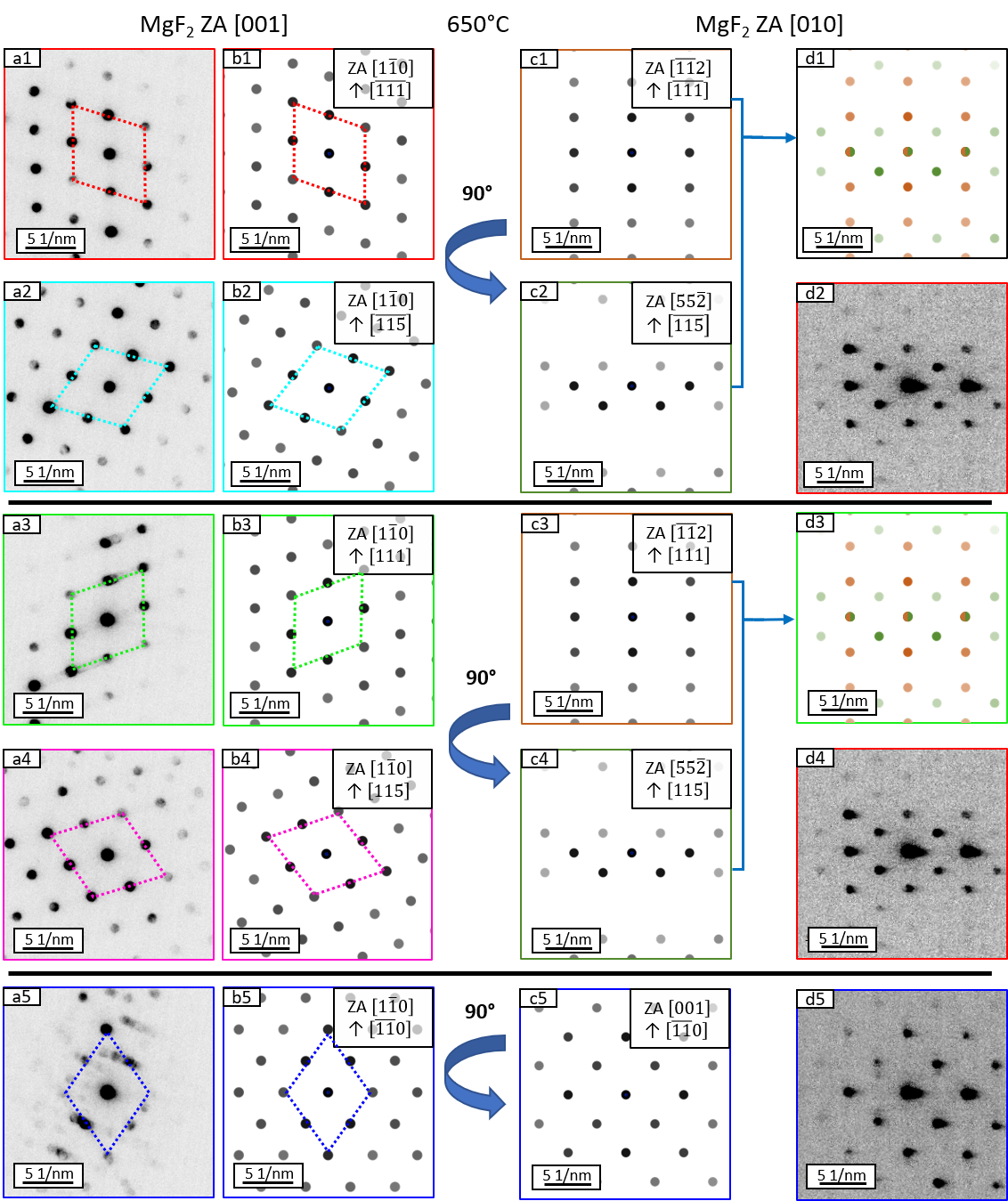}
\caption{Experimental NBED patterns of the 650\,$^{\circ}$C sample with their belonging simulated pattern. (a1)-(a5) are along the [001] ZA of the substrate and (d2), (d4) and (d5) are along the [010] ZA of MgF$_2$. The dashed rhombuses are a guide to the eye to identify the different orientations easier. The arrows indicate the change of the view direction from the [001] to the [010] zone axis which is a 90\,$^{\circ}$ rotation around the out of plane direction. (b1) and (b3) have the same pattern in the other direction as well as (b2) and (b4). As the orientations along the [001] ZA of MgF$_2$ are twins it is likely that the orientations are stacked along the other direction which leads to a superposition of the both diffraction patterns (c1) and (c2) same with (c3) and (c4). The color code is individual for each substrate orientation.}
\label{figureS2}
\end{figure}

As the mainly observed NBED patterns are the same for both temperatures they are only shown for the sample grown at 525\,$^{\circ}$C in the main text. In Fig.~\ref{figureS2} representative NBED patterns of the five mainly observed orientation of c-GaN are shown in the same way together with the calculated patterns as in Fig.~4 in the main text.

\begin{figure}[htp]
\centering
\includegraphics[width=.49\textwidth]{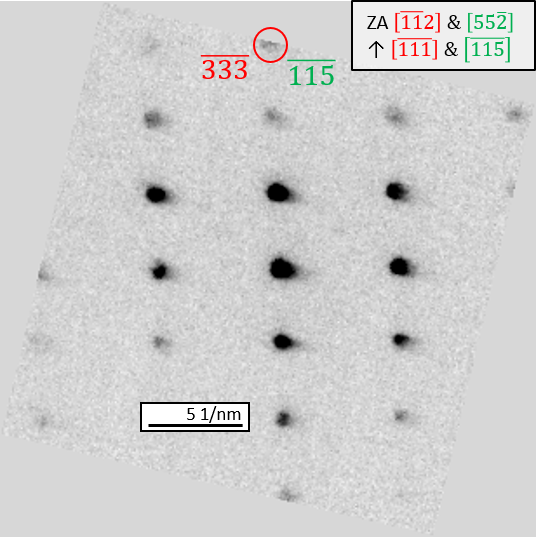}
\caption{NBED pattern of the 525\,$^{\circ}$C along the [010] zone axis of MgF$_2$. The pattern is rotated that the out of plane direction is vertical. At the top edge of the pattern the $\bar{3}\bar{3}\bar{3}/\bar{1}\bar{1}\bar{5}$ spot is marked. As this is a superposition of two diffraction patterns these spot can belong to both orientations.}
\label{figureS3}
\end{figure}

In Fig.~\ref{figureS3} a NBED pattern of the 525\,$^{\circ}$C sample is shown which is rotated to have the out of plane direction vertical. Instead to the other presented patterns the edges are not cut to show that we observe at the edge of this pattern the $\bar{3}\bar{3}\bar{3}/\bar{1}\bar{1}\bar{5}$ spot. As this pattern is a superposition of at least two c-GaN orientations having both a spot at these position.

\clearpage
\twocolumngrid 

\section{Mathematical treatment of twinning: cubic systems}\label{sec:SI2}

Cubic GaN crystallizes as the fcc zincblende structure. In such systems, twinning planes $\vec t$ are of type $\vec t =\{111\}$. There are several ways to describe twinning; here we choose a mirror operation at the twinning plane. For this purpose, we seek for the transformation matrix  $\matr T$ such that 
\begin{equation}
\label{eq2-01}
    \vec R = \matr T\cdot \vec r,
\end{equation}
where $\vec r$ and $\vec r'$ denote vectors in the untwinned and twinned crystal, respectively. 

In order to determine $\matr T$, we need to know the transformation of three linearly independent vectors, i.e. 
    \begin{align}
        \label{eq2-02}
        \vec r_i = (h_i,k_i,l_i) &\Rightarrow \vec r' = (h'_i,k'_i,l'_i) , i=1,2,3
    \end{align}

It is straightforward to show that this leads to a matrix equation
\begin{equation}
    \label{eq2-03}
    \matr A' = \matr T \cdot \matr A \mbox{\ \ or\ \ } \matr T = \matr A'\cdot \matr A^{-1}
\end{equation}
where
\begin{equation}
    \label{eq2-04}
    \matr A = \left( 
    \begin{array}{ccc}
         h_1 &h_2 &h_3  \\
         k_1 &k_2 &k_3 \\
         l_1 &l_2 &l_3
    \end{array}
    \right)
\end{equation}
and 
\begin{equation}
    \label{eq2-05}
    \matr A' = \left( 
    \begin{array}{ccc}
         h'_1 &h'_2 &h'_3  \\
         k'_1 &k'_2 &k'_3 \\
         l'_1 &l'_2 &l'_3
    \end{array}
    \right)
\end{equation}

There are different -- crystallographically equivalent -- ways to describe twinning in cubic systems, i.e. 
\begin{enumerate}
    \item Mirror operation at twinning plane (normal : $\vec t$). This way, we have $\vec t \rightarrow -\vec t$ and directions perpendicular to $\vec t$ are unaffected. Ex.: $\vec t = (111)$ will be transformed into $\vec t'= -(111)$, whereas $(1\overline 1 0) $ and $(\overline 1\overline 1 2)$ are unaffected. These considerations immediately show that successive twinning on $\{111\}$ planes containing the same $\langle 110\rangle$ direction -- as e.g. $(111)$ and $(11\overline 1)$ share [110] -- conserves this direction. Hence, in this case, untwinned crystal, primary twin and secondary twin share this $\langle 110\rangle$ direction, or related to electron diffraction, share this $\langle 110\rangle $ zone axis.
    \item Rotation about $\vec t$ by 180$^\circ (\pi)$. Same transformation except that all   directions are replaced by their opposite direction. 
    \item For $\vec t=\{111\}$ also the rotation about $\{110\}$ by $\arccos (1/3) \approx 70.52^\circ$ also leads to an equivalent transformation. 
\end{enumerate}
Here, we will follow route 1, i.e. mirror operation.  Hence, we have
    \begin{align}
        \label{eq2-06}
        \vec r_1&= \vec t = (h_1,k_1,l_1) \Rightarrow \vec r' = -\vec t = -(h_1,k_1,l_1) \nonumber \\
        \vec r_p&= (h_i,k_i,l_i) \Rightarrow \vec r_p' = \vec r_p = (h_i,k_i,l_i) , i=2,3\nonumber \\
        \vec r_p\cdot \vec t &=0
    \end{align}
    \subsection{Relevant cases for the c-GaN system}\label{sec21}
\subsubsection{Twinning on \texorpdfstring{$\vec t$}{t} = (111)}\label{sec211}
Following the above considerations, we obtain
 \begin{align}
        \label{eq2-06a}
        \vec t &= (111) \Rightarrow -\vec t = (\overline 1\overline 1\overline1) \nonumber \\
        \vec r_1&= (\overline 110) \Rightarrow \vec r_1' = \vec r_1 = (\overline 110) \nonumber \\
        \vec r_2&= (\overline 1\overline12) \Rightarrow \vec r_2' = \vec r_2 = (\overline 1\overline12) \nonumber 
    \end{align}
\begin{equation}
    \label{eq21-01}
    \matr A = \left( 
    \begin{array}{ccc}
         1 &\overline1 &\overline1  \\
         1 &1 &\overline 1 \\
         1 &0 &2
    \end{array}
    \right)
    \mbox{ and } \matr A' = \left( 
    \begin{array}{ccc}
         \overline 1 &\overline1 &\overline1  \\
         \overline1 &1 &\overline 1 \\
         \overline1 &0 &2
    \end{array}
    \right),
\end{equation}
which finally yields

\begin{equation}
    \label{eq21-02}
    \matr T_{111} = -\frac{1}{3}\left( 
    \begin{array}{ccc}
         \overline 1     &2             &2  \\
         2               &\overline 1   &2 \\
         2               &2             &\overline 1
    \end{array}
    \right).
\end{equation}    
    \subsubsection{Twinning on \texorpdfstring{$\vec t$ = $(\bar 1 11)$}{t = (-111)}}\label{sec212}
    An analogous treatment yields
    
\begin{equation}
    \label{eq22-01}
    \matr T_{\overline 111} = -\frac{1}{3}\left( 
    \begin{array}{ccc}
         \overline 1     &\overline 2             &\overline 2  \\
         2               &\overline 1   &2 \\
         \overline 2               &2             &\overline 1
    \end{array}
    \right).
\end{equation}  

    \subsubsection{Twinning on \texorpdfstring{$\vec t$ = $(1\bar 11)$}{t = (1-11)}}\label{sec213}
    An analogous treatment yields
    
\begin{equation}
    \label{eq23-01}
    \matr T_{1\overline 11} = -\frac{1}{3}\left( 
    \begin{array}{ccc}
         \overline 1     &\overline 2             & 2  \\
         \overline 2               &\overline 1   &\overline 2 \\
         2               &\overline 2             &\overline 1
    \end{array}
    \right).
\end{equation}  
    \subsubsection{Twinning on \texorpdfstring{$\vec t$ = $(11\bar 1)$}{t = (11-1)}}\label{sec214}
    An analogous treatment yields
    
\begin{equation}
    \label{eq24-01}
    \matr T_{11\overline 1} = -\frac{1}{3}\left( 
    \begin{array}{ccc}
         \overline 1     & 2             &\overline 2  \\
          2               &\overline 1   &\overline 2 \\
         \overline 2               &\overline 2             &\overline 1
    \end{array}
    \right).
\end{equation} 
\subsection{Secondary twins}\label{sec22}
As secondary twins, we denote the result of successive twinning operations on two (different) twin planes $j$ followed by $i$, i.e. the transformation
\begin{equation}\label{eq2a-01}
    \vec r'' = \matr T_i \vec r' = \matr T_i\matr T_j \vec r
\end{equation}
It is important to realize that for a single twin, we have 
\begin{equation}\label{eq2a-02}
    \matr T_i ^{-1} = \matr T_i,
\end{equation}
whereas for secondary twins, we have 
\begin{equation}\label{eq2a-03}
    \vec r = \matr T_j\matr T_i \vec r''.
\end{equation}

\subsection{Twin transforms of specific directions}\label{sec23}
Here, results of twin transforms of specific directions in a cubic system are summarized. It should be noted, that these calculations are \textit{independent} of any epitaxial relation to a substrate. For subsequent comparison to experimental data, however, certain projections need to be calculated.  Table~\ref{table01} summarizes results of two consecutive twin operations on different $\{111\}$ planes. As noted above, consecutive twinning on \textit{identical} twin planes will restore the untwinned orientation (compare also Table~\ref{table01}). 

In line with NBED data shown in Figs. 3 and 4 of the publication, consecutive twinning on $\{111\}$ planes sharing the same $\langle 110\rangle$ direction have to be considered. e.g. $(111)$ and $(11\overline1)$ sharing the $[1\overline10]$ direction. Inspection of the table (6th row) shows that [111] will be transformed parallel to [511] (twinning on $(11\overline1)$ and consecutive twinning on (111) transforms it parallel to $[\overline{11}\ \overline{11}\ 1]$, which is $\approx 3.68^\circ$ off $[\overline1\overline10]$ as observed in experiments. In the same way (not shown in Table~\ref{table01}), $[11\overline1]$ will be transformed parallel to $[11\overline5]$ (twinning on (111)) and further into $[\overline{11}\ \overline{11}\  \overline1]$, which is $\approx - 3.68^\circ$ off $[\overline1\overline10]$.  

Table~\ref{table01} further allows deduction of primary and secondary twinning on other $\{111\}$ planes, not sharing $1\overline10$. It turns out, that the considered directions mostly transform into high-indexed directions and in all cases, no additional diffraction spots will appear. Valuable insight is gained from the fact that $\{111\}$ diffraction spots can only be observed along $\langle 110 \rangle$ or $\langle 211\rangle$ zone axes. This allows for estimation of the amount of twinning on other $\{111\}$ planes by creating a virtual dark field by a (virtual) ring aperture containing $\{111\}$ spots. The dark regions of the resulting virtual dark field can be assumed to belong partly to the other $\{111\}$ planes.

%\addtolength{\tabcolsep}{-2.5pt}
\begin{table}[ht]
  \centering
  \caption{Table summarizing primary and secondary twin transformations of important lattice directions; for sake of clarity, transformed directions are indicated using integer Miller indices rather than providing fractional indices, i.e. only parallel directions are given. Please note, that the same transforms hold for lattice planes. PT: primary twin plane; ST: secondary twin plane, ST= (---) denotes cases without secondary twinning.}\label{table01}
  \begin{tabular}{|c|c|c|c|c|c|c|} \hline
    PT & ST & \ [001] \ & \ [1$\overline1$0] \ &\  [110] \ & \ [111] \ & \ [$\overline1\overline12 $]\ \\
    \hline
    (111) & (---) & $\overline 2\ \overline 2\ 1$ &  1\ $\overline1$\ 0 & $\overline1\ \overline1\ \overline4$ & $\overline1\ \overline1\ \overline1$ & $\overline1\ \overline1\ 2 $\\
    & (111) & $0\ 0\ 1$ & $1\ \overline1\ 0$ & $1\ 1\ 0$ & $1\ 1\ 1$ & $\overline1\ \overline1\ 2$\\
    & $(\overline111)$ & $\overline4\  \overline8\  1$ & $\overline1\ 1\ 4$ & $\overline{11}\ 5\  \overline4$ & $\overline5\ \overline1\ \overline1$ & $1\ \overline7\ 2$\\
    & $(1\overline11)$ & $\overline8\ \overline4\ 1$ & $\overline1\ \overline1\ \overline4$ & $5\  \overline{11}\  4\  $ & $\overline1\ \overline5\ \overline1$ & $\overline7\ 1\ 2$\\
    & $(11\overline1)$ & $4\ 4\ \overline7$ & $1\ \overline1\ 0$ & $\overline7\ \overline7\ \overline8$ & $\overline1\ \overline1\ \overline5$ & $5\ 5\ \overline2$\\
    \hline
    $(\overline 111)$ & (---) & $2\ \overline2\ 1$ & $\overline1\ 1\ 4$ & $1\ 1\ 0$ & $5\ 1\ 1$ & $1\ \overline7\ 2$\\
     & $(111)$ & $4\ \overline8\ 1 $ & $\overline{11}\ 5\ 4$ & $\overline1\ \overline1\ \overline4$ & $1\ \overline{11}\ \overline{11}$ & $11\  \overline{13}\  14$\\
     & $(\overline111)$ & $0\ 0\ 1$ & $1\ \overline1\ 0$ & $1\ 1\ 0$ & $1\ 1\ 1$ & $\overline1\ \overline1\ 2$\\
     & $(1\overline11)$ & $\overline4\ 4\ \overline7$ & $\overline7\ 7\ 8$ & $1\ 1\ 0$ & $5\ 13\ \overline7$ & $\overline{17}\ \overline1\ \overline{14}$\\
     & $(1 1 \overline1)$ & $8\ \overline4\ 1$ & $5\ 11\ \overline4$ & $\overline1\ \overline1\ 4$\ & $5\ \overline7\ 13$ & $19\ \overline5\ \overline{10}$\\
    \hline
    $(1\overline11)$ & (---) & $\overline2\ 2\ 1$ & $\overline1\ 1\ \overline4$ & $1\ 1\ 0$ & $1\ 5\ 1$ & $\overline7\ 1\ 2$\\
     & $(111)$ & $\overline8\ 4\ 1$ & $5\ 11\ \overline4$ & $\overline1\ \overline1\ \overline4$ & $\overline{11}\ 1\ \overline{11}$ & $\overline{13}\ 11\ 14$\\
     & $(\overline111)$ & $4\ \overline4\ \overline7$ & $\overline7\ 7\ \overline8$ & $1\ 1\ 0$ & $13\ 5\ \overline7$ & $\overline1\ \overline{17}\ \overline{14}$\\
     & $(1\overline11)$ & $0\ 0\ 1$ & $1\ \overline1\ 0$ & $1\ 1\ 0$ & $1\ 1\ 1$ & $\overline1\ \overline1\ 2$\\
     & $(11\overline1)$ & $\overline4\ 8\ 1$ & $\overline{11}\ \overline5\ \overline4$ & $\ \overline1\ \overline1\ 4$ &\ $\overline7\ 5\ 13$ &\ $\overline5\ 19\ \overline{10}$\\
     \hline
    $(11\overline 1)$ & (---) & $2\ 2\ 1$ & $1\ \overline1\ 0$ & $\overline1\ \overline1\ 4$ & $\ 5\ 1\ 1$ & $1\ \overline7\ 2$\\
     & $(111)$ & $\overline4\ \overline4\ \overline7$\ & $1\ \overline1\ 0$\ & $\overline7\ \overline7\ 8$ & $\overline{11}\ \overline{11}\ 1$ & $\overline1\ \overline1\ \overline{22}$\\
     & $(\overline111)$ & $8\ 4\ 1$ & $\overline1\ 1\ 4$ & $5\ \overline{11}\ 4$ & $13\ \overline7\ 5$ & $11\ 19\ \overline2$\\
     & $(1\overline11)$ & $4\ 8\ 1$ & $\overline1\ 1\ \overline4$ & $\overline{11}\ 5\ 4$ & $\overline7\ 13\ 5$ & $19\ 11\ \overline2$\\
     & $(11\overline1)$ & $0\ 0\ 1$ & $1\ \overline1\ 0$ & $1\ 1\ 0$ & $1\ 1\ 1$ & $\overline1\ \overline1\ 2$\\
     \hline\hline
  \end{tabular}
\end{table}

\subsection{Reciprocal space maps (RSM) and electron diffraction patterns}\label{sec24}
Up to this point, only twinning transforms have been discussed without referring to special epitaxial relations and cross-section directions used for obtaining electron diffraction patterns or X-ray RSM. Experimentally, cross-sections along MgF$_2$ [010] and [001] have been analyzed revealing two main orientational relationships: 
GaN~(110) $\parallel$ MgF$_2$~(100) with GaN~$[1\overline10]$ $\parallel$ MgF$_2$~[001] (in short: $[1\overline10](110)$~c-GaN $\parallel$ [001](100)~MgF$_2$) and GaN~(111) $\parallel$ MgF$_2$~(100) with GaN~$[1\overline10]$ $\parallel$ MgF$_2$~[001] (in short: \\ $[1\overline10](111)$c-GaN $\parallel$ [001](100)MgF$_2$) and their respective variants as a result of the twofold symmetry of the substrate surface (compare also Table~\ref{table02}). For the former (subsequently denoted as '2/2-epitaxy'), this variant is fully equivalent due to the twofold symmetry of the layer along [110]. For the latter (subsequently denoted as '2/3-epitaxy'), the variant corresponds to a twinning transform with (111)~c-GaN as a twinning plane, hence the notation as '2/$\overline3$-epitaxy'. 

\begin{table*}[ht]
\caption{Epitaxial relations observed in this work for c-GaN growth on (100) MgF$_2$. The lattice mismatch $\delta$ has been calculated via $\delta = 1-d/D, $ where $d$ and $D$ are lattice plane spacings in GaN and MgF$_2$, respectively.   The common growth directions in MgF$_2$ and GaN are  denoted as UVW  and uvw,  respectively. $^\dag$ Note, that there is small misfit ($\approx 0.1\%$) comparing 5$\times$($\overline1\overline12$)-GaN and 4$\times$(020)-MgF$_2$. \label{table02}}
\begin{center}
\begin{tabular}{|l|ccl|ccl|r|}
\hline
\multirow{2}{*}{phase / epitaxy} & \multicolumn{3}{c|}{ MgF$_2$} & %
    \multicolumn{3}{c|}{ GaN } & \multirow{2}{*}{$\delta$ [\%]}\\
\cline{2-7}
& UVW & HKL & D [nm] & uvw & hkl & d [nm] & \\
\hline
\multirow{2}{*}{c-GaN / 2/2 epitaxy} & \multirow{2}{*}{100} & 020& 0.2310 & \multirow{2}{*}{110}  & 002 & 0.2260 & 2.2\\
& & 002 & 0.1525 & & 2$\Bar{2}$0  & 0.1598 & -4.8 \\
\hline
\multirow{2}{*}{c-GaN / 2/3 epitaxy } & \multirow{2}{*}{100} & 020 & 0.2310 & \multirow{2}{*}{111}  &  $\overline1\overline12$ & 0.1846 & 20.0$^\dag$\\
& & 002 & 0.1525 & & $2\overline20$  & 0.1598 & -4.8 \\
\hline
\multirow{2}{*}{c-GaN / 2/$\overline3$ epitaxy } & \multirow{2}{*}{100} & 020 & 0.2310 & \multirow{2}{*}{$\overline1\overline1\overline1$}  &  $\overline1\overline12$ & 0.1846 & 20.0$^\dag$\\
& & 002 & 0.1525 & & $2\overline20$  & 0.1598 & -4.8 \\
\hline
\end{tabular}
\end{center}
\end{table*}

\begin{figure*}
    \centering
    \includegraphics[width=0.8\textwidth]{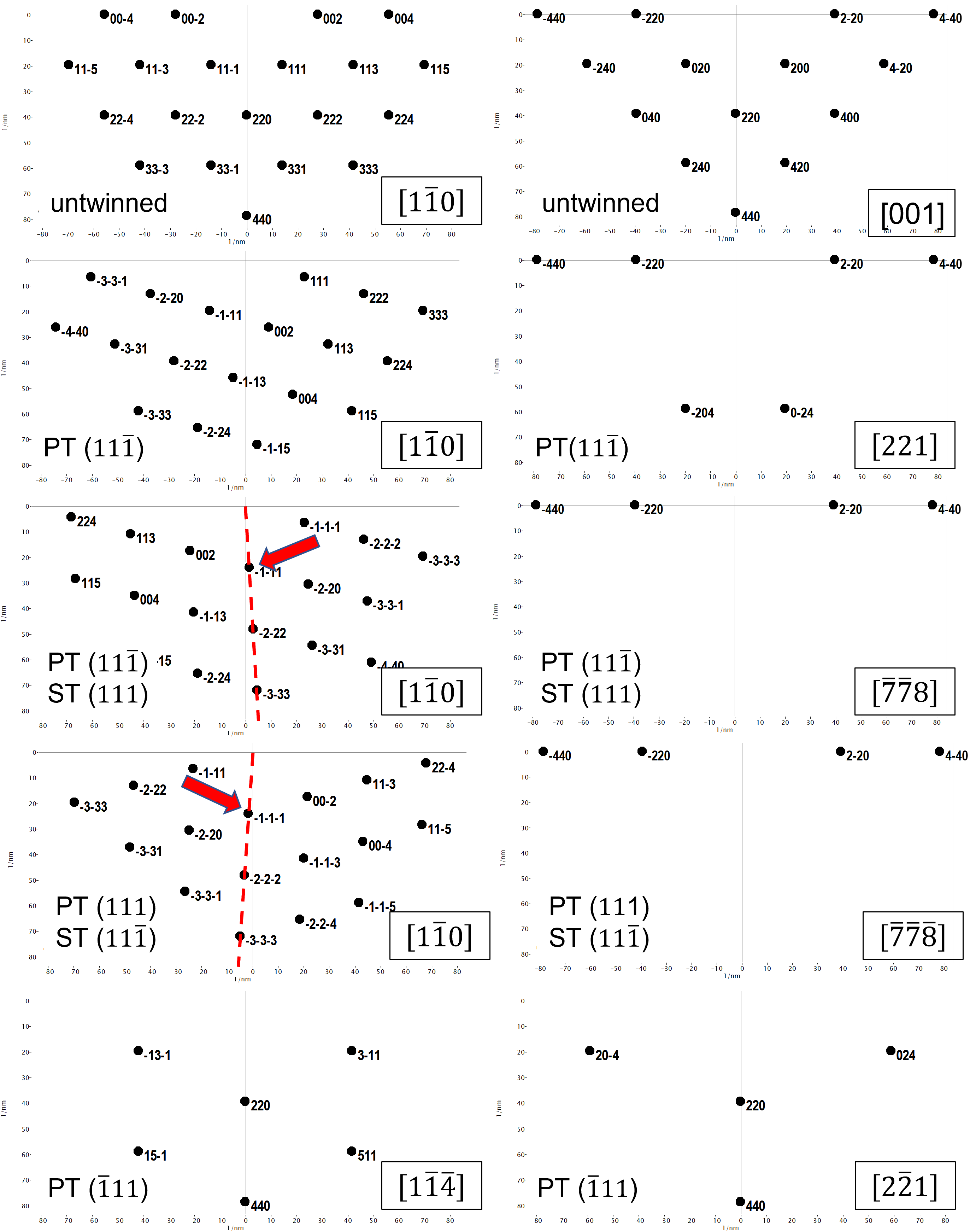}
    \caption{Electron diffraction patterns for the 2/2-epitaxy for cross-sections along MgF$_2$~[001] (left column) and along MgF$_2$~[010] (right column); PT: primary twin plane, ST: secondary twin plane. GaN zone axes are indicated in boxes in the lower right of each pattern.  First row: untwinned. Second row: primary twin on $(11\overline1)$ plane leading to GaN~$[\overline1\overline15]$ parallel to the main growth direction.  Third row: primary twin on $(11\overline1)$ plane and secondary twin on (111); please note, that in this case, GaN~$[\overline1\overline11]$ is 3.68$^\circ$ off the main growth direction. Fourth row:primary twin on $(111)$ plane and secondary twin on $(11\overline1)$. Fifth row: primary twin on $(\overline111)$. Dashed lines indicate a tilt of $\{111\}$ direction by $\pm 3.68^\circ$ from the normal resulting from secondary twins; arrows indicate the related positions of $\overline1\overline11$ and $\overline1\overline1\overline1$ diffraction spots.  }
    \label{figS4}
\end{figure*}

Subsequently, electron diffraction patterns (or equivalently RSM) are presented for the two systems in order to show possible spot overlap and additional spots possibly specific for certain variants. Fig.~\ref{figS4} summarizes specific diffraction patterns for the 2/2-epitaxy. Comparing the different patterns, the result summarized in Table~\ref{table01} are visualized. An important observation is the tilt of $(\overline1\overline11)$ and $(\overline1\overline1\overline1)$ planes by $\approx \pm3.68^\circ$, which explains the splitting of (111) spots observed by X-ray diffraction (compare Fig.8 in the manuscript), which averages over many twin domains, i.e. effectively superimposes individual patterns. 

\begin{figure*}
    \centering
    \includegraphics[width=0.8\textwidth]{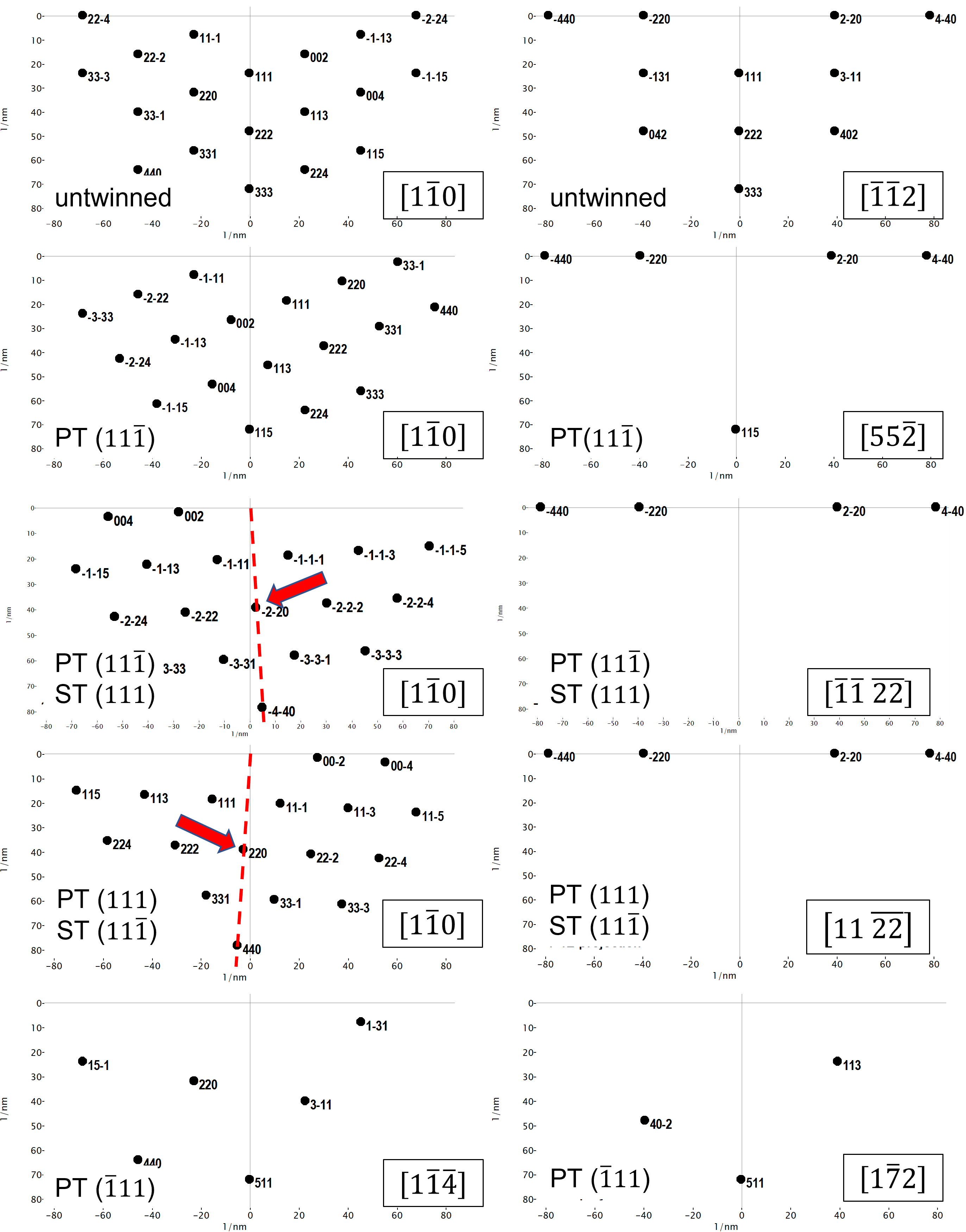}
    \caption{Electron diffraction patterns for the 2/3-epitaxy for cross-sections along MgF$_2$~[001] (left column) and along MgF$_2$~[010] (right column); PT: primary twin plane, ST: secondary twin plane. GaN zone axes are indicated in boxes in the lower right of each pattern.  First row: untwinned. Second row: primary twin on $(11\overline1)$ plane leading to GaN~[115] parallel to the main growth direction.  Third row primary twin on $(11\overline1)$ plane and secondary twin on (111). Fourth row: the analogous operation for 2/$3$-epitaxy.  Fifth row: primary twin on $(\overline111)$. Dashed lines indicate a tilt of $\{220\}$ direction by $\pm 3.68^\circ$ from the normal resulting from secondary twins; arrows indicate the related positions of $220$ and $\overline2\overline20$ diffraction spots. }
    \label{figS5}
\end{figure*}

Fig.~\ref{figS5} summarizes specific diffraction patterns for the 2/3-epitaxy; note, that patterns due to 2/$3$-epitaxy are simply obtained by $180^\circ$ rotation about the Q$_z$ axis (compare e.g. rows 3 and four). Comparing the different patterns, the result summarized in Table~\ref{table01} are visualized. An important observation is the tilt of $(\overline2\overline20)$ and $(220)$ planes by $\approx \pm3.68^\circ$, which should give rise to splitting of (220) spots in X-ray diffraction reciprocal space maps.  

